\documentclass[11pt,a4paper]{article}
\pdfoutput=1
\usepackage{jcappub}

\usepackage[T1]{fontenc} % Needed
\usepackage{graphicx}
\usepackage{adjustbox}  % For the table of the priors
\usepackage{booktabs}
\usepackage[T1]{fontenc}
\usepackage[utf8]{inputenc}
\usepackage{url}
\usepackage{appendix}
\usepackage{amsmath}
\usepackage{mathtools}
\usepackage{subfig}
\usepackage{epstopdf}
\usepackage{listings}
\usepackage{verbatim}   % useful for program listings 
\usepackage{color}      % use if color is used in text
\usepackage{tabularx,ragged2e}
\usepackage{booktabs}

\def\bea{\begin{eqnarray}}
\def\eea{\end{eqnarray}}
\def\be{\begin{equation}}
\def\ee{\end{equation}}
\def\fr{\frac}

\def\la{\label}
\def\be{\begin{equation}}
\def\ee{\end{equation}}
\def\ci{\cite}

\def\ksm{km\,s$^{-1}$Mpc$^{-1}$\,}
\def\H{H$_0$\;}
\def\sH{\sigma_H}

\def\le{\left}
\def\ri{\right}

\def\L{\Lambda}
\def\ac{a_c}
\def\LCDMd{$\Lambda$CDM\;}
\def\LCDM{$\Lambda$CDM}
\def\LCDMxd{$\Lambda$CDM-Nx\;}
\def\LCDMx{$\Lambda$CDM-Nx}

\def\SHES{SH$_0$S}

\def\ad{a_{\star}}
\def\rex{\rho_{ex}}
\def\Oex{\Omega_{ex}}
\def\H{H$_0$\;}

\title{\boldmath Towards a Solution to the H$_0$ Tension:  the Price to Pay}

\author[a]{Axel de la Macorra\note{Corresponding author},}
\author[b]{ Erick Almaraz}
\author[a]{and Joanna Garrido}
\affiliation[a]{Instituto de F\'isica, Universidad Nacional Aut\'onoma de M\'exico, \\ Circuito de la Investigaci\'on Cient\'ifica Ciudad Universitaria, 04510  CDMX, M\'exico }
\affiliation[b]{African Institute for Mathematical Sciences AIMS, Cape Town, South Africa }
\emailAdd{macorra@fisica.unam.mx} \emailAdd{ealmaraz@gmail.com} \emailAdd{joannagarr@ciencias.unam.mx}

\abstract{The tension between  current expansion rate H$_0$ using Planck data and direct model-independent measurements in the local universe has reached  a tension above 5$\sigma$ in the context of the $\Lambda$CDM model. The growing tension among early time and local  measurements of  H$_0$  has not ameliorated and remains a crucial and open question in  cosmology. Solutions to understand  this tension are possible hidden sources of systematic error in the observable measurements or modifications to the  concordance  $\Lambda$CDM  model. In this work, we investigate a solution to the H$_0$ tension by modifying $\Lambda$CDM  and we add at  early times  extra relativistic energy density $\rho_{ex}$   beyond the Standard Model.  For scale factor larger than $a_c$ this extra energy density $\rho_{ex}$ dilutes faster than radiation and becomes subdominant. In some context this $\rho_{ex}$  corresponds to Early Dark Energy (EDE) or  by Bound Dark Energy (BDE)  and we refer to this cosmological model as $\Lambda$CDM-Nx. We implement $\Lambda$CDM-Nx   in CAMB and perform a full COSMO-MC (MCMC)  allowing to simultaneously fit the latest data from CMB anisotropies and the value of H$_0 = 74.03\pm 1.42,\mathrm{km\, s^{-1}Mpc^{-1}}$ obtained by distance ladder measurements using Cepheid variables to calibrate the absolute luminosity of Type Ia supernovae  by  A. Riess [R-19] \ci{Riess:2019cxk}. The inclusion of $\rex$ ameliorates the tension between early and late time  measurements only slightly  and we obtain a value H$_0 = (68.70\pm 0.45$\ksm)  still in conflict with local measurements [R-19].   We follow up our analysis by proposing two forecasting standard deviation  $\sigma_H=1$ and $\sigma_H=0.5$  (in units of \ksm)  for local  distance measurements, i.e. H$_0 = (74.03\pm 1)$ \ksm  and H$_0 = (74.03\pm 0.5$) \ksm.  We implement these new values of H$_0$ in MCMC and we  obtain  a value of  H$_0 =(72.83\pm 0.47)$ \ksm\,  at 68\% confidence level  for $\sigma_H =0.5$,  fully consistent with  [R-19],  while the price to pay is a  percentage increase of  $0.12\%$  in  CMB $\chi_{cmb}^2$. Finally,  the extra  energy density $\rho_{ex}$  leaves distinctive imprints in the matter power spectrum   at scales $k \sim k_c$ with  $k_c=a_cH(a_c)$  and in the CMB power spectrum, allowing for  independent verification of our analysis.}


\begin{document}
\maketitle
\flushbottom

\section{Overview}\label{sec:intro}

Even before the discovery of the accelerated expansion rate of the Universe \cite{Roukema:1993yra, Riess:1998cb, Perlmutter:1998np}, the quest for determining the rate of expansion of the Universe  has occupied a central role in cosmology for decades. 
However,  great technical and observational achievements in recent years have delivered percentage-level precision measurements of the cosmological parameters. 
The improvement probing the physics of different epochs of the universe has yielded discordance in some measurements. In particular  we have an increasing tension among the value of some cosmological parameters as for example the value of  H$_0$.

According  to the distance ladder measurements,  which uses  close by  Cepheids  to anchor supernovae data and determine their distance \citep{Riess:2016jrr, Riess:2018byc, Riess:2019cxk},   the obtained value for the Hubble parameter is up to $3.4\sigma$ higher than the value determined using Cosmic Microwave Backgound (CMB) probes. This distance ladder method to determine H$_0$ is model-independent, i.e. it does not rely on  the underlying cosmological model,  and the latest estimated reports a value  of H$_0 = 74.03\pm 1.42$ \ksm  \cite{Riess:2019cxk}.  

An alternative calibration of the distance ladder uses the tip of the red giant branch (TRGB) method \cite{Beaton:2016nsw}.  This method is independent of the Cepheid distance scale and gives a value of H$_0 = 69.8 \pm 1.9$\, \ksm \cite{Freedman:2019jwv} which is midway in the range defined by the current Hubble tension. It agrees at the $1.2\sigma$ level with that of the Planck collaboration \cite{Aghanim:2018eyx}, and at the $1.7 \sigma$ level with the SH$_0$ES  (Supernovae H$_0$ for the Equation of State) measurement of $H_0$ based on the Cepheid distance scale
\cite{Riess:2019cxk}.
Measurements of lensing time delays \cite{Bonvin:2016crt, Birrer:2018vtm, Wong:2019kwg}  between multiple images of background quasars provide a high value for H$_0$ which is in agreement with the traditional local distance ladder estimation. In   \cite{Bonvin:2016crt} they report a value of  H$_0 =71.9^{+2.4}_{-3.0}$ \ksm, while \cite{Birrer:2018vtm} finds a value H$_0 = 72.5^{+2.1}_{-2.3}$ \ksm, and the H0LiCOW (H$_0$ Lenses in CosmoGrail Wellspring) team \cite{Wong:2019kwg}, reports H$_0 = 73.3^{+1.7}_{ -1.8}$ \ksm,  by making a joint analysis of six gravitationally lensed quasars with measured time delays. This technique is completely independent of both the supernovae and CMB analyses. 

The precise CMB measurements  by Planck [P-18]  determines  a value of  the Hubble parameter as \H = ($67.27 \pm 0.60$)\ksm at 68\% confidence level assuming  the standard \LCDMd model,  corresponding  the particle content of the standard model of particles physics \ci{pdg}  supplemented by cold dark matter and  a cosmological constant as dark energy.   The value of \H  from CMB is  in conflict with  the value of \H determined at late cosmological times  from local measurements in Riess et al  [R-19] \cite{Riess:2019cxk}  given by  \H= ($74.03 \pm 1.42$)\ksm  and with an average reported value  \H=($73.3 \pm 0.8$)\ksm\, \ci{Verde:2019ivm}  from different local measurements projects.   
A combined analysis of distance measurements for four megamaser-hosting galaxies done by the Megamaser Cosmology Project (MCP) \cite{Pesce:2020xfe} reports a value H$_0=73.9\pm 3.0$ \ksm.  A combination of time-delay cosmography and the distance ladder results is in 5.3$\sigma$ tension with Planck CMB determinations of \H in flat LCDM.

On the other hand, the latest results from the Planck collaboration  report a model dependent \LCDM\, extrapolated value of H$_0 = (67.36 \pm 0.54)$ \ksm \cite{Aghanim:2018eyx}, while the latest results the Atacama Cosmology Telescope \cite{Aiola:2020azj} CMB probe found a value that agrees with the Planck satellite estimate within $0.3\%$, reporting a value of H$_0 = 67.6\pm 1.5$ \ksm.  In a recent review  L.Verde, A. Riess and T.True  reported an average value of local measurements  of  \H=$73.03\pm 0.8$ \ksm \cite{Verde:2019ivm}. Regardless of the exact value of \H by local measurements, the significance of the tension between the measurements of early and late time  lies in the range  4.0$\sigma$ and 5.7$\sigma$ \cite{Verde:2019ivm}, implying some profound miss-understanding in either the systematic errors of the observational analysis, or the theoretical \LCDM\, model.  

The discrepancy between the early and late-Universe \H measurements has gained major attention by the cosmological community  and some authors have explored a variety of extensions to the minimal \LCDM\, model to accommodate the high value of \H obtained by local measurements with the precise information encoded in the CMB. 
Trying to understand the tension between these two values led to a reexamination of possible sources of systematic errors in the observations \cite{Addison:2015wyg}, \cite{Aylor:2018drw},  \cite{Efstathiou:2013via}, but it also suggests the need to extend our physical model describing the universe.
Any of these theoretical modifications should leave  the accurate determination of the angular scale of the acoustic peaks in the CMB power spectrum by Planck unchanged \citep{Knox2019}.

The \H\, tension has been studied recently in  \cite{ Poulin:2018cxd, Karwal:2016vyq, Sakstein:2019fmf,Knox:2019rjx, DiValentino:2019ffd}  and more recently in  \cite{Freese:2021rjq,Murgia:2020ryi, Niedermann:2020dwg,DiValentino:2021izs, DiValentino:2020zio,Hill:2020osr, Klypin:2020tud,2020PhRvD.102j3502I}
where the impact on structure formation has been studied.

The suggestion made by \cite{Riess:2016jrr} to explore the existence of dark radiation in the early Universe in the range of $\Delta N_{eff} = 0.4-1$ to solve this tension was explored in detail by \cite{Bernal:2016gxb} where they explore changing the value of  $N_{eff}$ and $c_s$.
Alternatively, some models explore the possibility of having interactions between the dark sector that can, not only help to solve the cosmic coincidence problem but also solve the \H tension \cite{DiValentino:2019jae, Pan2019}. In the logic of exploring alternative models for the dark sector, in \cite{DiValentino:2020naf} investigated the possible scenarios for a phantom crossing dark energy component as another option for solving the Hubble tension.
Given the amount of interest invested in this topic, some authors have explored changes to General Relativity in order to accommodate the high value of H$_0$ with CMB  data. For instance, the work by  \citep{Gomez-Valent:2020mqn} explores a model in which a fifth force between dark matter particles is mediated by a scalar field which plays the role of dark energy. In another work, models which vary the effective gravitational constant and effective number of relativistic degrees of freedom are explored by \citep{Ballesteros:2020sik}. In a different approach, \citep{Kreisch:2019yzn} explores the possibility of strongly interacting massive neutrinos to alleviate the \H tension.

However, perhaps the most widely explored  extension to \LCDMd is known as Early Dark Energy (EDE) \cite{Linder:2008nq, Francis:2008md, Calabrese:2010uf,Appleby:2013upa}. There is no  unique or unambiguous  definition of EDE. Typically  in EDE models there is an early period  during which an extra  energy component, not contained in \LCDM\, model, contributes to the expansion rate of the universe H.  Even the terminology of  "early period"  is model and case dependent as it can take place in radiation domination era or at late times as for example at $z\sim 4$.  
Original  Early Dark Energy  models where  motivated by the evolution of scalar fields (quintessence) to describe the evolution of Dark Energy \cite{Riess:2006fw, Doran:2006kp,Linder:2008nq, Francis:2008md, Bean:2001wt}. This EDE models  had in general a none negligible  energy density at early times  well in radiation domination. The   equation of state  $w$ of the quintessence scalar field  had a period of $w=1/3$ at early times  with a later transition  to  $w\sim 1$, diluting the energy density and becoming subdominant for a long  period  of time covering most of the matter domination  period,  to finally reappear dynamically at late time as dark energy  and were originally studied in \cite{Steinhardt:1999nw,Zlatev:1998tr} and \cite{delaMacorra:1999ff,delaMacorra:2004mp,delaMacorra:2001ay,delaMacorra:2001xx} and  \cite{Caldwell:2005tm,Linder:2007wa}.  
Alternatively, recent  EDE  models add an extra component to the energy-momentum tensor  $\Omega_{\textrm{EDE}}(a)$ at different scales and this EDE  dilutes rapidly at scale factor $a_c$, which determines the time of the transition, with  $\Omega_{\textrm{EDE}}=0$  for $a \gg a_c$. This EDE modifies the expansion rate of the universe, the cosmological distances and  the density perturbations at different epochs   \cite{Calabrese:2011hg,Calabrese:2010uf} and  \cite{Samsing:2012qx, Poulin:2018cxd,Keeley:2019esp} and have been proposed as  deviations from \LCDMd and  possible solutions to \H crisis \cite{Riess:2019cxk}.

Furthermore, the increasing statistical tension in the estimated Hubble parameter from early and late times observations \cite{Verde:2019ivm} has reignited interest in alternative cosmological models, while the surge in  clustering data \cite{Alam:2020sor} and the percentage precision for cosmic distances \cite{Alam:2020sor,Aghanim:2018eyx} allows to search for extensions beyond $\Lambda$CDM by  searching for cosmological features in the matter \cite{Gomez-Navarro:2020fef,Klypin:2020tud,Knox:2019rjx,Francis:2008md, Calabrese:2011hg,Calabrese:2010uf,Samsing:2012qx}
or CMB power spectra,  standard distances rulers, or tensions in \LCDMd model as the recent \H crisis \cite{Riess:2019cxk}.

On the other hand a  physically motivated Dark Energy  model  presented in \cite{delaMacorra:2004mp, delaMacorra:2018zbk,Almaraz:2018fhb,Almaraz:2019zxy} introduces a dark sector, corresponding to a dark gauge group SU(3) similar to the strong QCD interaction in the standard model.  The fundamental particles contained in this dark SU(3) are massless and redshift as radiation for $a<\ac$ but  the underling dynamics of the gauge interaction of this group forms massive bound states once the interaction becomes strong,  similar as with  protons and neutrons in the strong QCD force, and we refer to this model as "Bound Dark Energy model" (BDE) \cite{delaMacorra:2004mp, delaMacorra:2018zbk, Almaraz:2018fhb}. The energy  of the elementary particles is transferred  to the lightest  bound state  after the phase transition takes place at $a_c$ and  corresponds to  a scalar field $\phi$.  Due to the dynamics of  $\phi$ the energy density of BDE dilutes at $\ac$ and eventually reappears close to present time as Dark Energy \cite{delaMacorra:2018zbk,Almaraz:2018fhb}.  This dilution at $\ac$ leaves interesting imprints on the matter power spectrum  \cite{Almaraz:2019zxy} (for a model independent analysis see for instance \cite{Jaber:2019opg, delaMacorra:2020zqv}). 
BDE is a particular model of elementary particles physics where an extra gauge group $SU(3)$ is introduced and contains naturally  the main characteristics of EDE, namely  it  accounts for extra  relativistic  energy density $\rex$  at high energies  while $\rex(a)$ dilutes rapidly  for $a>a_c$  due to a phase transition of the underlying gauge  and forms bounds states \cite{delaMacorra:2018zbk,Almaraz:2018fhb}.

The  main goal in this work  is study the tension and possible solution  in the value of H$_0$ from low redshift probes with the precise determination of CMB data. 
This paper is organised as follows. Section \ref{sec:a_scale}  we present  a brief introduction and the details behind our modifications through a toy model calculations in section  \ref{subsec:toymodel}.   The working details  and  implementation in the Boltzmann code CAMB and COSMOMC  are presented in section \ref{sec.results}, the results are discussed  in section \ref{sec.results} and  the analysis are in section \ref{sec.analysis}, while we present our conclusions in \ref{sec:conclusions}.

\section{Introduction} \label{sec:a_scale}

The  main goal in this work  s study the tension  and possible solutions  in the value of H$_0$ from low redshift probes and the precise determination of CMB data. 
We work with two different cosmological models, the first one is simply the standard  \LCDM\, model, corresponding to a content of the standard model of particles physics \ci{pdg}, cold dark matter and a cosmological constant as dark energy, while the second model  we denote as  \LCDMx,  corresponding to \LCDM\,  but
supplemented with  extra relativistic energy density $\rex (a) sim 1/a^3$ present at a scale factor  $a\leq a_c$, while for  $a> \ac$ $\rex(a)$ dilutes as $\rex(a) \sim 1/a^6$. This model  \LCDMxd is  inspired by BDE \ci{delaMacorra:2018zbk, Almaraz:2018fhb}.

For definiteness in this study we take the recent local measurement  $H_0=(74.03 \pm 1.42)$ \ksm\,  at 68\% confidence level from Riess et al [R-19] \cite{Riess:2019cxk}, and  the  inferred value of  H$_0=(67.27\pm 0.60)$ \ksm  at  one-$\sigma$ level   from Planck-2018  [Pl-18] \cite{Aghanim:2018eyx} for a \LCDMd model  using   (TT,TE,EE+LowE)  measurements.
We  modified   CAMB \cite{Lewis:1999bs, Howlett:2012mh}  and make a full 
 \texttt{COSMO-MC} (Markov Chains)  analysis  \footnote{http://cosmologist.info/cosmomc} \cite{Lewis:2013hha,Lewis:2002ah,Lewis:2019xzd}. We perform the analysis for \LCDM\, and \LCDM-Nx models. However,   besides the one-sigma value $\sigma_H=1.42$  from local \H\, measurements [R-19]  and we also consider 
two forecasting one-sigma values  $\sigma_H$, and for definiteness we choose and we introduce in the analysis a value of  $H_0=(74.03 \pm \sigma_H $ \ksm\,
with $\sigma_H =1$  and $\sigma_H=0.5$  (in units of \ksm). With these two  forecasting values   we asses the impact of a more precise local \H  measurements on the posterior  value of \H\,   from CMB + local \H  data.  Notice however, that our forecasting  one-sigmas $\sH=1$ and $\sH=0.5$  are  similar  to  the one reported in  \cite{Verde:2019ivm}
with  \H=$(73.03\pm 0.8)$\ksm.  The results of our analysis are shown in section \ref{sec.analysis}  and we present the conclusions in (\ref{sec:conclusions}).

The value of  \H=$74.03\pm1.42$\ksm \,  measurements determined by A. Riess and his team [R-19] \cite{Riess:2016jrr, Riess:2018byc, Riess:2019cxk} has a discrepancy  between  4.0$\sigma$ and 5.8$\sigma$ \cite{Verde:2019ivm}  with the Planck's CMB   (TT,TE,EE+LowE) \cite{Aghanim:2018eyx} [P-18]  inferred value of  H$_0=(67.27\pm 0.60)$ \ksm  at  one-$\sigma$ level  in a  \LCDMd model.  The solution  to this discrepancy remains an open question in cosmology. Since CMB  radiation is generated at an early  epoch  $\ad=1/1090$, the prediction of the Hubble constant at present time \H\, inferred by Planck data is a consequence of the assumption of the validity  of  the standard \LCDM\, model.
So either Planck or local \H\, measurements  are inaccurate, due to possible systematics, or we need to modify  the concordance  cosmological  \LCDM\, model.  Here we follow this second option and  we will attempt  to reconcile  the value of \H of these two observational experiments.

We will work with two different cosmological models. The first one is simply the standard  \LCDM\, model, corresponding to a content of the standard model (sm) particles physics \ci{pdg}  and a cosmological constant as dark energy.  We name the second model as \LCDMx\, and it consists of  \LCDM\, supplemented by an extra relativistic  energy density $\rex (a)\sim 1/a^3 $  present at early times for a scale factor  $a\leq a_c$, where $a_c$ denotes the transition scale factor, and we  assume that $\rex(a)\sim 1/a^6$ for $a > a_c$, motivated by BDE and EDE models. 

We have implemented  the   cosmological  \LCDMxd\,  model  in  CAMB \cite{Lewis:1999bs, Howlett:2012mh}  and 
and we perform a full  COSMO-MC (Markov Chains)  analysis for several data sets  described  in section ({sec.results})   for both  \LCDM\, and \LCDM-Nx models 
and we present  the results and conclusions in section (\ref{sec.results}).

For definiteness we take  the   (TT,TE,EE+lowE)  measurements  from Planck-2018   [Pl-18] \cite{Aghanim:2018eyx}  with H$_0=(67.27\pm 0.60)$ \ksm 
and the recent local measurement  $H_0=(74.03 \pm 1.42)$ \ksm\,  at 68\% confidence level from Riess et al [R-19] \cite{Riess:2019cxk}.
However, besides the one-sigma value $\sigma_H=1.42$  (in units of \ksm) from local measurements [R-19],  we also introduce two forecasting one-sigma values and we choose $\sigma_H=1$  and $\sigma_H=0.5$. With these two  forecasting values   we want to asses the impact a more precise local \H  measurements on the posterior  value of \H\,   combined with  the same CMB as before.  Notice however, that our forecasting  one-sigmas values $\sH=1$ and $\sH=0.5$  are  of the same order as  in the average value   \H=($73.03\pm 0.8$)\ksm reported in  \cite{Verde:2019ivm}.  The results of these analysis are shown in section \ref{sec.analysis}  and we present the conclusions in (\ref{sec:conclusions}).
Before discussing the results  of the implemented  the   cosmological  models \LCDM\, and \LCDM-Nx models  in  CAMB and COSMO-MC   presented in section \ref{sec.results} we would like to follow up a simple toy model presented in in section \ref{subsec:toymodel}  illustrating  how an extra relativistic energy density $\rex(a)$,  present only at early times  $a<a_c$,  can account for the same acoustic scale $\theta (\ad)$ as  measured by Planck [P-18] but with the value of \H consistent  with  Riess [R-19]. We estimate the cosmological constraints analytically in section \ref{sec.Oex} and we  study the impact of the extra  $\rex$  in the  growth of the linear matter density and in the matter power spectrum in section \ref{sec.PS}.

\section{Cosmological Toy Models}\label{subsec:toymodel}

We present now a simple toy model illustrating  analytically how adding an extra relativistic energy density $\rex(a)$, present at early times,
can account for having the same acoustic scale $\theta(\ad)$ as \LCDMd model  but with a higher value of H$_0$.

 \subsection{Acoustic Scale}\label{sec.acoustic}

Planck satellite  \cite{Aghanim:2018eyx} has delivered impressive quality cosmological data by measuring the CMB background radiation.  Perhaps the most accurate measurements is the acoustic scale anisotropies  given by the acoustic angle $\theta$ defined in as the ratio  of the comoving sound horizon  $r_s(\ad)$  and  the comoving angular diameter distance  $D_A(\ad)$ evaluated at recombination scale factor $\ad $   (with a  readshift $z_{_\star} =1/\ad -1 \simeq 1089 $)  as
\be 
\theta (\ad)=\fr{ r_s(\ad)}{D_A(\ad)}.
\la{theta}\ee
The  (TT,TE,EE+lowE) CMB Planck-2018   \cite{Aghanim:2018eyx}
 measurements at 68\%  confidence level gives 
\be
100\,\theta(\ad) =(1.04109\pm 0.0003),
\label{th} 
\ee
in the context of the standard \LCDM\, model, corresponding to a flat universe with cold dark matter (CDM),  a cosmological constant $\L$ as dark energy and the Standard Model particles \ci{pdg}.
The coomoving angular diameter distance and the acoustic scale are defined as 
\be 
D_A(\ad)=\int^{a_o}_{\ad} \fr{da}{a^2H(a)}, \;\;\;\;\;\;\;\;\;\;\; r_s(\ad)=\int^{\ad}_{a_i}  \fr{c_s}{a^2H(a)}\; da
\la{da}\ee
with $H(a)\equiv \dot a/a$ the Hubble parameter and $c_s$  the sound speed,
\be
c_s(a)=\fr{1}{\sqrt{3(1+R)} }, \;\;\;\;\;\;\;\;\;\;\; R \equiv  \fr{3}{4}  \fr{\rho_b}{\rho_\gamma} = \fr{3}{4}  \le(\fr{\Omega_{bo}}{\Omega_{\gamma o}}\ri)\le(\fr{a}{a_o}\ri).
\ee
Since Planck  CMB measurements determines the angle  acoustic $\theta(\ad) =r_s(\ad)/D_A(\ad)$ accurately,  any modification of  \LCDM\, must clearly preserved the ratio  in  $\theta(\ad)$.  
A  larger value of \H  reduces $D_A(\ad)$ and $r_s(\ad)$,
however since the integrations limits  differ in $D_A(\ad)$ and $r_s(\ad)$ a change in H$_0$  will modify the angle  $\theta(\ad)$.

Let us take two models,   the standard  \LCDMd model  (or "sm")  and   \LCDMxd (also referred as  "smx")  corresponding to a \LCDMd  with additional relativistic particles for $a<\ad$. Imposing the constraint  to have  the same acoustic scale  $\theta(\ad)$ in these two models, i.e.
 \be
\theta(\ad)=\fr{r_s^{sm}(\ad)}{D_A^{sm}(\ad)}= \fr{r_s^{smx}(\ad)}{D_A^{smx}(\ad)},
\la{rdrd}\ee
the relative quotient of $r_s(\ad)$  and $D_A(\ad)$  of these models must satisfy
\be
\xi  \equiv \fr{D_A^{smx} (\ad)}{D_A^{sm}(\ad)} = \fr{r_s^{smx}(\ad)}{r_s^{sm}(\ad)}.
\la{ddrr}\ee
Any change in $D_A(\ad)^{smx}/D_A^{sm}(\ad)$,  due for example  for a different amount of H$_0$,  can be compensated with a change in $r_s(\ad)^{smx}/r_s(\ad)^{sm}$
to maintain the same  $\theta(\ad)$.

We impose the constraint to  have the  same acoustic scale $\theta(\ad)$ in both models, with   \LCDM\,  (i.e. "sm") having a value of  \H as measured by  Planck-2018 [P18],  where we take for presentation purposes  H$_0^P=67$ (in units of \ksm),  and the second model \LCDMxd (i.e. "smx"), corresponding
to the  standard  \LCDM\, model with  extra  relativistic energy density $\rex(a)$  and an \H  given by  H$_0^R=74$  (in units of \ksm),   consistent awith A. Riess et al  [R19].
We define the Hubble parameter in \LCDM\, as $H_{sm}^2=(8\pi G/3)\rho_{sm}$  with an energy content
$\rho_{sm}= \rho_r^{sm}+\rho_m^{sm}+\rho_\Lambda^{sm}$ for radiation, matter  and  cosmological constant, respectively, while 
\LCDMx\,  has  $H_{smx}^2=(8\pi G/3)\rho_{smx}$  with $\rho_{smx}\equiv \rho_r^{smx}+\rho_m^{smx}+\rho_\Lambda^{smx}$.
For simplicity we assume  the same amount of matter in both models and we take  for model $\rho^{smx}$  the following content: \\
i) for   $a \leq a_c  $  we have extra radiation  $\rho_{ex}\neq 0$ with $\rho_r^{smx}=\rho_r^{sm}+ \rex$ and $ \rho_m^{smx}=\rho_m^{sm}$\\
ii) for $a>a_c$,  we have $\rho_{ex}=0$,   $\rho_r^{smx}=\rho_r^{sm}$, $ \rho_m^{smx}=\rho_m^{sm}$ but  $\rho_\L^{smx}  > \rho_\L^{sm}$.

We will now determine the relation between the amount $\rex(a_c)$ (or equivalently $\Oex (a_c)$) as a function of  the transition scale $a_c$  such that the ratio of the sound horizon  $r_s(\ad)$ at decoupling and the angular distance to the last scattering surface $D_A(\ad)$ is unchanged, preserving thus the acoustic angle $\theta (\ad)$  as measured by Planck  \cite{Riess:2019cxk},   but with a Hubble parameter \H  in \LCDMxd\, ("smx") model consistent with the high value of local measurements   H$_0^R=74$ \cite{Riess:2019cxk}. 
Taking  $H_0^{sm}=H_0^P=67$ and  $H_0^{smx}=H_0^R=74$  the ratio $D_A^{smx} (\ad)/D_A^{sm}(\ad)$  gives 
\be
\xi = D_A^{smx} (\ad)/D_A^{sm}(\ad)=0.981. 
\la{DAA}\ee
Since  $\xi <1$  and using eq.(\ref{ddrr})  we  require $r_s^{smx}(\ad) /r_s^{sm} (\ad)$ to be smaller than one. We can achieve this by
increasing H(a) in the region $a\leq \ad$  in \LCDMxd compared to the standard \LCDM\, model by introducing  extra radiation $\rex(a)$ in the region $a<\ad$.  With this  modification we  tune $\rex(a_c)$ to obtain the ratio  $r_s^{smx}(\ad)/r_s^{sm}(\ad)$  to obtain  the same value  of $\theta(\ad)$  in eq.(\ref{ddrr})
as measured by Planck-2018.

Let us compare the Hubble parameter in these two models in the region $a\ll  a_o$, where dark energy is subdominant, giving
\be
\fr{H_{sm}}{H_{smx}}=\sqrt{ \fr{\rho_r^{sm}+\rho_m^{sm}}{\rho_r^{sm}+\rho_m^{sm}+ \rex}}=\sqrt{1 -\Oex}
\la{HH}\ee
with 
\be
\Omega_{ex}  \equiv  \fr{\rho_{ex}}{\rho_{smx}}=  \fr{\rho_{ex}}{\rho_{sm}+\rex}\simeq \fr{N_{ex}\,\beta}{1+(N_\nu+N_{ex})\,\beta}
\la{Oex}\ee
and the last term in eq.(\ref{Oex}) is given in terms of relativistic degrees of freedom  with  $ \rho_{sm}=g_{sm}\rho_\gamma$,  $\rho_{ex}=g_{ex}\rho_\gamma $ and  $g_{sm}  =  1+ N_\nu\beta$,\,  $g_{ex} = N_{ex}\beta$, \,$g_{smx}=g_{sm}+g_{ex}$ with $\rho_{\gamma}=\fr{\pi^2}{30}\; g_{\gamma} T^4_{\gamma}$,  $N_\nu=3.046$   and $N_{ex}$ the extra relativistic degrees of freedom in terms of the neutrino temperature and $\beta = \le(7/8\ri)  \le(4/11\ri)^{4/3}$.
Notice that last approximation in eq.(\ref{Oex}) is only valid  in radiation domination epoch.
Since we assume in our toy models the same amount of matter and radiation in models  \LCDM\,   and \LCDMx\,  at present time but different values of H$_0$,  we must necessarily have  a larger amount of dark energy in model  \LCDMx\, than in \LCDMd  to account for the increase value  in H$_0$.
We constrain  \LCDMx\,  model  by imposing that it gives  same  acoustic angle $\theta (\ad)= r_s(\ad)/D_A(\ad)$ as  \LCDM 
(c.f. eq.(\ref{rdrd}))  and  relative quotient of $r_s(\ad)$  and $D_A(\ad)$ as in eq.(\ref{ddrr}).

% \subsubsection{Hubble parameter and Dark Energy}\la{sec.HH}

We will now compare the Hubble parameter H in  \LCDM\,(sm) and \LCDMx\,(smx) models. By our working hypothesis both models have the same amount of matter and radiation
at present time, while the value of \H  differs with H$_0^P=67$ for \LCDM\,(sm) and H$_0^R=73$ for \LCDMxd (smx). 
Let us express $H$ as
\be
H^{sm}(a)=H_{0}^{sm}\sqrt{\Omega_{mo}^{sm}(a/a_o)^{-3}+\Omega^{sm}_{ro}(a/a_o)^{-4} +\Omega^{sm}_{\Lambda o}}
\la{hsm}\ee
and
\be
H^{smx}(a)=H_{0}^{smx}\sqrt{\Omega_{mo}^{smx}(a/a_o)^{-3}+\Omega^{smx}_{ro}(a/a_o)^{-4} +\Omega^{smx}_{\Lambda o}}
\la{hsmx}\ee
with the constraint $\Omega_{mo}+\Omega _{ro}+\Omega_{\Lambda o}=1$ for  both models. Since by assumption we have  the same amount of matter and radiation,
$\rho_{mo}^{sm}=\rho_{mo}^{smx}$ and  $\rho_{ro}^{sm}=\rho_{ro}^{smx}$,  we simply multiply and divide by the critical density $\rho_{co}$ of each model
to get
\bea
\rho_{q o} = \Omega_{q o}^{sm}\,\rho_{co}^{sm}= \Omega_{qo}^{smx}\,\rho_{co}^{smx},\;\;\;\;\;\;
\rho_{\Lambda o }^{smx}   =\rho_{\Lambda o }^{sm}   +  \rho_{c o}^{sm} \le ( \fr{(H_0^{smx})^2}{ (H_0^{sm})^2} -1 \ri)
\la{rmro}\eea
with $q=m,r$ for matter and radiation, respectively.  Models  \LCDM\, (sm) and \LCDMx\, (smx) have the same $\rho_{mo}$ and $\rho_{ro}$ but a different amount of  $H_0$ 
gives  a different amount of dark energy  $\rho_{\Lambda }$ as seen in eq. (\ref{rmro}).
We clearly see  in eq.(\ref{rmro}) how  different amounts of $H_0$ 
impacts the  Dark Energy density  in these two models. We further express 
\be
\Omega_{q o}^{smx}=\fr{(H_0^{sm})^2}{(H_0^{smx})^2}\;\Omega_{qo}^{sm},  \;\;\;\;\;\;\;\;\;\;\;\;
\Omega_{\Lambda o}^{smx}  =1 - (\Omega_{mo}^{smx}+\Omega_{ro}^{smx})=1 - \fr{(H_0^{sm})^2}{(H_0^{smx})^2} \le( \Omega_{mo}^{sm}+ \Omega_{ro}^{sm}\ri)
\la{omor}\ee
The Hubble parameter $H$  becomes
\be
H^{s} (a) =H_{0}^{s} \sqrt{1+  \Omega_{mo}^{s}  \le[(a/a_o)^{-3} -1\ri]+\Omega^{s} _{ro} \le[(a/a_o)^{-4}-1\ri] }
\la{ha}\ee
with $s=sm, smx$ for \LCDMd and \LCDMx\, models, respectively.  Expressng  $H^{smx}$ in terms of $sm$ quantities  we have for \LCDMx,
\bea
H^{smx} (a) &=&H_{0}^{smx}  \sqrt{1+ \fr{(H_0^{sm} )^2}{(H_0^{smx} )^2}\le(  \Omega_{mo}^{sm}  \le[(a/a_o)^{-3} -1\ri]+\Omega^{sm} _{ro} \le[(a/a_o)^{-4}-1\ri] \ri) } \\
&=& H_{0}^{sm}   \sqrt{ \fr{(H_0^{smx} )^2}{(H_0^{sm}  )^2}  + \Omega_{mo}^{sm} \le[(a/a_o)^{-3} -1\ri]+\Omega^{sm} _{ro} \le[(a/a_o)^{-4}-1\ri] }.
\la{hb}\eea
We have expressed $H^{smx}(a)$ in terms of quantities of model $sm $ and the ratio $H_0^{sm} /H_{0}^{smx} $.  
 The difference in $H^{sm}$ in \LCDM\, and $H^{smx}$ in \LCDMx\,  due to the distinct values of $H_0$  is manifested
 in the first terms in the square root in eqs.(\ref{ha}) and (\ref{hb})  ($"1"$ in eq.(\ref{ha})  compared to $(H_0^{smx} /H_0^{sm} )^2$ in eq.(\ref{hb}))
with $(H_0^{smx} /H_0^{sm}  )^2=(H_0^{R}/H_0^P)^2=(74/67)^2=1.22$ for our  two fiducial examples.

\subsection{Impact of $\rex$ on the Acoustic Scale $r_s(a_{\star})$ }\la{sec.Oex}

We will  now quantify  the  impact on the acoustic scale $r_s(\ad)$ from extra relativistic energy  density $\rex(a)$,   present before recombination,  
helps to conciliate the  $H_0$ tension between early and late time measurements.   
 We assume that  $\rex$ is present up to the scale factor $a_c$ and than it dilutes rapidly \ci{delaMacorra:2020zqv} and no longer contributes to $H$.
 This rapid dilution of $\rex$ can be motivated by  Bound Dark Energy model \cite{delaMacorra:2018zbk,Almaraz:2018fhb}  or by EDE models \cite{Poulin:2018cxd, Sakstein:2019fmf}. Interestingly, the rapid dilution of $\rex$ besides contributing towards a solution to the $H_0$ crisis  may also leave interesting signatures in the matter power spectrum \cite{delaMacorra:2020zqv,Almaraz:2018fhb, Jaber:2019opg, Keeley2019, Gomez-Navarro:2020fef}  which can be correlated with the $H_0$ solution. 
 
\begin{figure*}[ht]
\begin{center} 
%\begin{multicols}
\includegraphics[height =0.22\textheight]{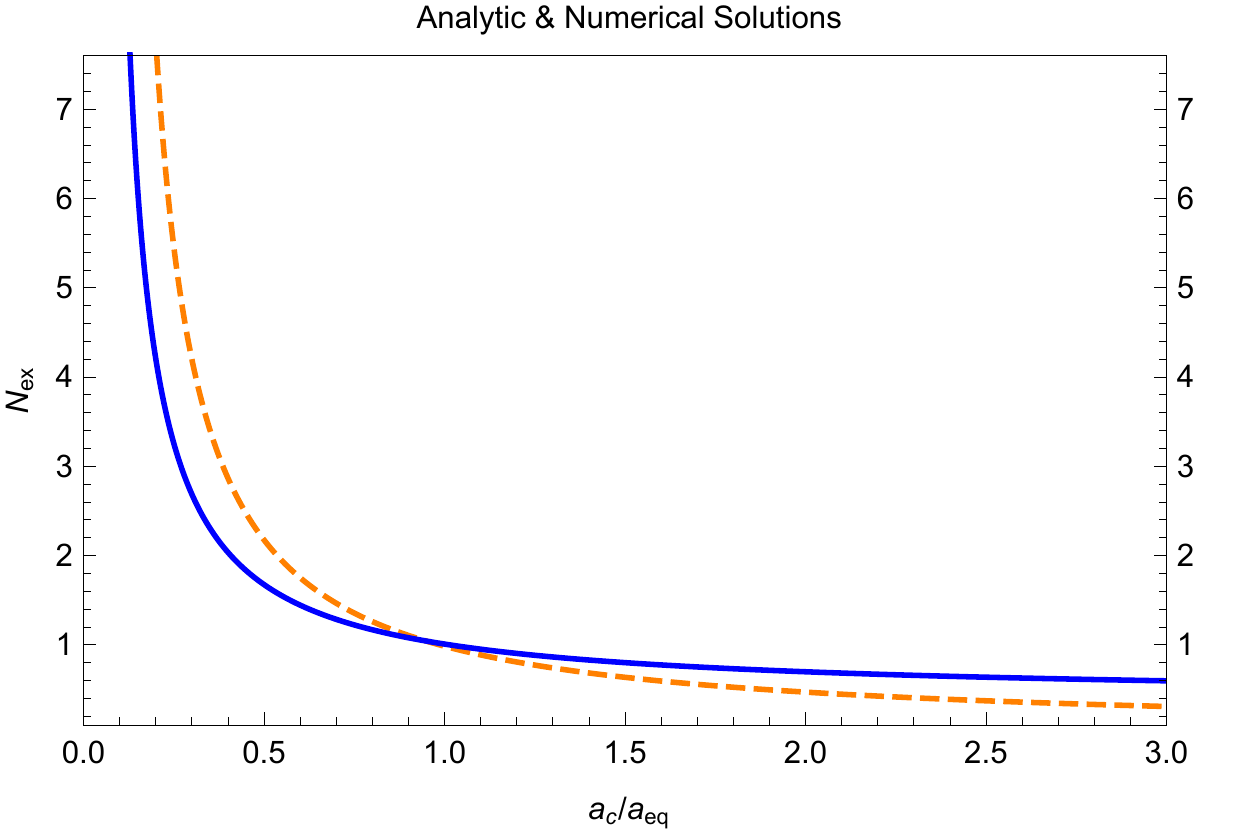}     
\includegraphics[height =0.22\textheight]{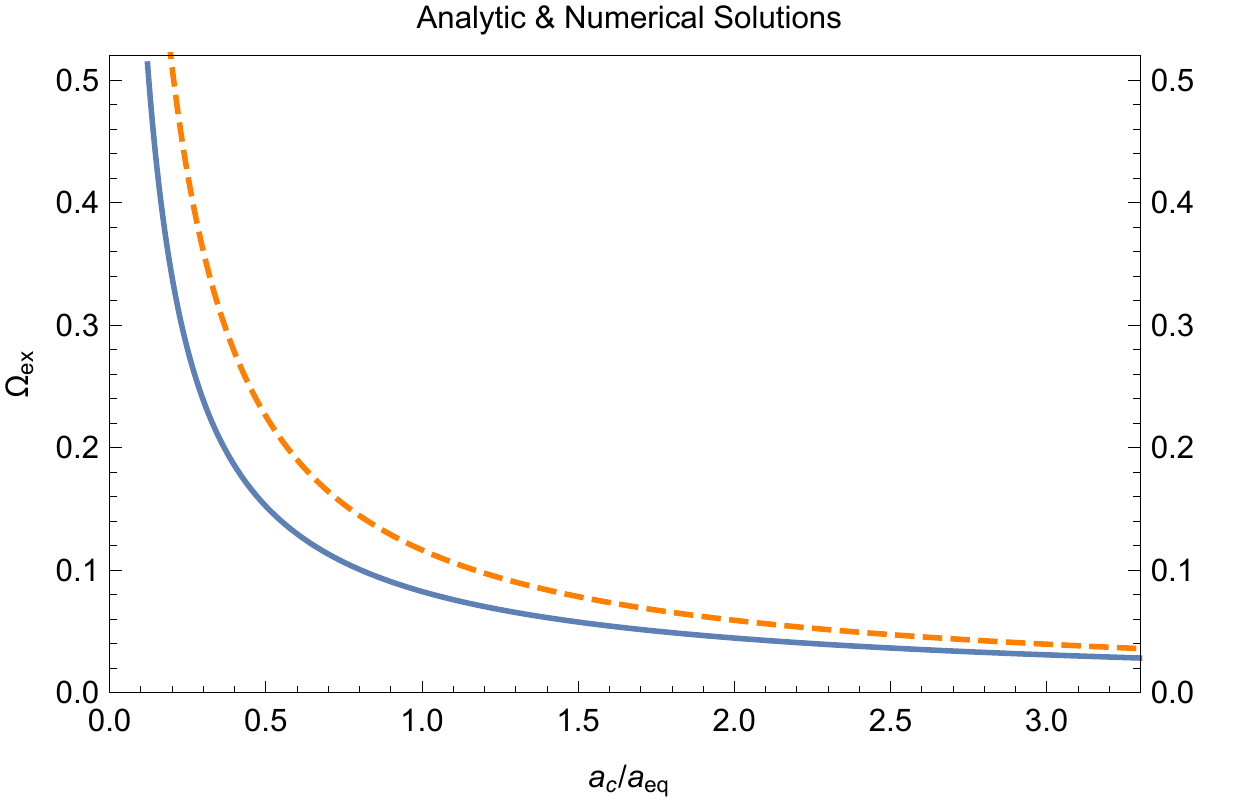}
%\end{multicols}
\caption{ Analytic and Numerical Solutions $N_{ex}(a_c)$ and $\Oex(a_c)$. We plot  $N_{ex}$  (left panel) and  $\Oex (a_c)$ (right panel)  as a function of $x\equiv a_c/a_{eq}$   satisfying  the constraint  $ r_s^{smx}(\ad)/r_s^{sm}(ad)=\xi$ (c.f.  eq.(\ref{DAA})). Numerical solution (blue)  and analytic solution from eq.(\ref{adac2}) (dashed-orange). }
\label{fignex2}
\end{center} 
\end{figure*}

Let us now  study the  impact of $\rex$ on the $H_0$ tension  problem and  its cosmological signatures.
From eq.(\ref{HH}), we take   $H_{sm}/H_{smx }=\sqrt{1-\Oex}$  and for simplicity and presentation purposes  we consider  $\Oex$ constant  for $a\leq \ac$  and    $\Oex=0$, $H_{smx}=H_{sm}$   for $a >\ac$. 
The precise  impact of $\Oex$ and the value of $a_c$ in the different cosmological parameters must be numerically calculated and a full implementation in a
Boltzmann code Markov Chains (here we use CAMB and COSMO-MC \cite{Lewis:1999bs, Howlett:2012mh, Lewis:2013hha,Lewis:2002ah,Lewis:2019xzd})  is  presented in section (\ref{sec.results}).
Nevertheless having an approximated  analytic expressions  of the acoustic scale  allow us to have  a simple grasp  of the impact of $\rex$ and $a_c$ in the magnitude of
 $r_s(\ad)$ and in the possible solution to the $H_0$ crisis.
The  change in the acoustic scale $r_s(\ad)$  in models \LCDMd ($sm$) and \LCDMxd ($smx$)  can be easily estimated. 
Let us consider the difference
\bea
 r_s^{sm}(\ad) -r_s^{smx} (\ad)  &=&  \int^{\ad}_{a_i} \fr{c_s da}{a^2H_{sm}}-  \int^{\ad}_{a_i} \fr{c_s da}{a^2H_{smx}} \\
&=& \int^{a_c}_{a_i} \fr{c_s da}{a^2H_{sm}}-  \int^{a_c}_{a_i} \fr{c_s da}{a^2H_{smx}} \equiv  r_s^{sm}(a_c)- r_s^{smx}(a_c)
\la{rsss}
\eea
where  we have taken into  account that $H_{smx}=H_{sm}$ for $a > a_c$ and the integrals from $a_c\leq a\leq \ad$ cancel out.  
As long as  $H_{sm}/H_{smx}=\sqrt{1-\Oex}  $  is constant   we can  simply express 
\be
r_s^{smx}(\ac) \equiv   \int^{\ac}_{a_i} \fr{c_s\, da}{a^2H_{smx}}=\int^{\ac}_{a_i} \le(\fr{H_{sm}}{H_{smx}}\ri)\fr{c_s\, da}{a^2H_{sm}} =  \sqrt{1-\Oex} \;\;r_s^{sm}(a_c).
\la{rsac}\ee
Clearly  the amount of  $\Oex$ determines the ratio of $r_s^{smx}(\ad)/r_s^{sm}(\ad) $.
Now writing $ r_s^{sm}(\ad) - r_s^{smx} (\ad) = r_s^{sm}(\ad)(1 - \xi)$  with $\xi$ given in  eq.(\ref{ddrr})    and  $ r_s^{sm}(a_c) - r_s^{smx} (a_c) = r_s^{sm}(a_c)(1 - \sqrt{1-\Oex} )$
from eq.(\ref{rsss}) we obtain $r_s^{sm}(\ad)(1-\xi)=r_s^{sm}(a_c)(1 - \sqrt{1-\Oex} )$ and
\be
\fr{r_s^{sm}(\ad)}{r_s^{sm}(a_c)} = \fr{1-\sqrt{1-\Oex}}{1- \xi  }.
\la{drrr}\ee
If we assume  radiation domination, the quantity $a^2H$ is constant,  and taking for simplicity and presentation purposes $c_s$  also constant we get 
\be
\fr{r_s^{sm} (\ad)} {r_s^{sm} (a_c)}=\fr{\int^{\ad}_{a_i} 
 \fr{ da\; c_s}{a^2 H_{sm}}}{\int^{a_c}_{a_i}  \fr{ da\; c_s}{a^2H_{sm}}}=\fr{a_c H_{sm}(a_c) }{ \ad H_{sm}(\ad) } =\fr{\ad}{a_c}
\la{rsrcc}\ee
and eq.(\ref{drrr}) becomes
\be
 \le(\fr{\ad} {a_c}\ri)  = \fr{1-\sqrt{1-\Oex}}{ 1- \xi } = 52.63\,  (1-\sqrt{1-\Oex})
\la{adac2}\ee
where we have set  $\xi=0.981$, our fiducial value in eq.(\ref{ddrr}). We obtain in eq.(\ref{adac2})  a very simple analytic solution  for  $a_c$ as a function
of $\Oex$ with the the constraint to have the same acoustic angle $\theta(\ad)$ (c.f. eq.(\ref{theta})) in \LCDMx\, with H$_0=74$ as in \LCDM\, model with H$_0=67$.
We see in eq.(\ref{adac2})  that   larger   values of  $a_c$  require  smaller amount of $\Oex$. We plot in fig.(\ref{fignex2}) the required value of $N_{ex}$ and $\Oex$ as function of $x = a_c/a_{eq}$ using eq.(\ref{adac2})  and from the numerical calculation solving the full $H(a)$ as given in eqs.(\ref{hsm}) and (\ref{hsmx}).
We should keep in mind that eq.(\ref{adac2})  is only an approximation since we assumed radiation domination, 
however it gives a simple estimation of the amount of  $\Oex$  as a function of  $a_c$ required to accommodate a consistent model with same acoustic scale with Planck data [P-18] and Riess \H value [R-19].

\subsection{Matter Power Spectrum  and  $\rex$ }\la{sec.PS}

We have seen in the previous section how $\rex $  impacts cosmological distances and contributes to reduce the \H\, tension.  Let us now studythe effect of  the rapid dilution of $\rex$ not only  affects  the evolution of density perturbations and in the matter power spectrum $P(k,z)$  \cite{delaMacorra:2020zqv},\cite{Gomez-Navarro:2020fef} and  \cite{Klypin:2020tud,Knox:2019rjx,Francis:2008md, Calabrese:2011hg,Calabrese:2010uf,Samsing:2012qx}.
Interesting, an  energy density $\rex(a)$ that dilutes rapidly at $a=a_c$ (see section (\ref{sec.Oex}) will leave detectable imprints on the matter power spectrum  which can be correlated with a possible  solution to H$_0$ tension.
We can estimate the location and magnitude of this bump produced at the transition scale $a_c$  corresponding to the mode
\be
k_c\equiv \ac H_c
\la{kc}\ee
with $H_c\equiv H(a_c)^2=(8\pi G/3)\rho_{smx}(a_c)$ and  $\rho_{smx}=\rho_{sm}+\rex$.
The amplitude of the bump  is related to the magnitude of $\rex(a)$  while the width of the bump is related to how fast $\rex$ dilutes \cite{delaMacorra:2020zqv}. 
In radiation domination the amplitude  $\delta_m=\delta \rho_m/\rho_m$ has a logarithmic growth 
\bea
\delta_m^{smx} (a)  &=& \delta_{mi}^{smx}  \le(\textrm{ln}(a/a_h^{smx})+1/2 \ri), \\
\la{dmL}\delta_m^{sm} (a)  &=& \delta_{mi}^{sm}  \le(\textrm{ln}(a/a_h^{sm})+1/2 \ri)
\eea
where $a_h$ corresponds to horizon crossing.
Comparing this growth for the same mode $k^{smx}=k^{sm}$ with $k^{sm}=a_h^{sm} H^{sm} (a_h^{sm})$  and  $k^{smx}=a_h^{smx} H^{smx}(a_h^{smx})$. 
Modes $k>k_c$ cross the horizon at $a_h < a_c$ and we find from eq.(\ref{HH})
\be
\fr{a_h^{smx}}{a_h^{sm}}= \fr{H^{sm}}{H^{smx}} =\sqrt{1-\Oex}.
\ee
The ratio $\Delta \delta_m=\delta_m^{smx}/\delta_m^{sm}=(\delta_{mi}^{smx}/\delta_{mi}^{sm})  (\textrm{ln}(a/a_h^{smx})+1/2)/(\textrm{ln}(a/a_h^{sm})+1/2)$ can be expressed for $a>\ac$ as
\be
\Delta \delta_m =  \fr{ \delta_{mi}^{smx}}{\delta_{mi}^{sm}}\;
 \fr {\le[  \le(H_+^{smx}/H_-^{smx}\ri) \textrm{ln} \le(a/a_c\ri) +\textrm{ln} \le( a_h^{sm}/a_h^{smx}\ri) + \textrm{ln} \le(a_c/a_h^{sm}\ri)+\fr{1}{2}   \ri] }
{  \textrm{ln}\le(a/a_c\ri) + \textrm{ln}\le(a_c/a_h^{sm}\ri) +\fr{1}{2} },
\la{DD}\ee
where  $H_+^{smx}(a_c)$ contains $\rex$ and $H_-^{smx}(a_c)$  has $\rex=0$. 
For presentation purposes here we have consider a step function at $a_c$ with $\rex(a)=0$ for $a<a_c$ and we have
$H_+^{smx}(a_c)/H_-^{smx}(a_c) = H^{smx}(a_c)/H^{sm}=1/\sqrt{1-\Oex}$.
Eq.(\ref{DD}) is valid for modes  $k>k_c$. entering the horizon at  $a_h < a_c$.
The increase for modes $k >k_c$ at present time is 
\be
\Delta \delta_m= \fr{\delta_m^{smx}}{\delta_m^{sm}} =  \fr{\delta_{mi}^{smx}}{\delta_{mi}^{sm}} \fr{H_{+}^{smx}}{H_{-}^{sm}}
= \fr{\delta_{mi}^{smx}}{\delta_{mi}^{sm}} \fr{1}{\sqrt{1-\Oex}}
\ee
where we assumed for $a_o\gg \ac$.  On the other hand modes $k<k_c$  do not undergo the transition  and are not boosted by
the rapid dilution of $\rex$. The final result in the matter power spectrum is the generation of a  bump in the ratio
$P_{smx}/P_{sm}$ at scales of the order of $k_c$.

To conclude,  we have seen in our toy model that an extra relativistic energy density $\rex$  may alleviate the tension in the \H\, measurements and at the same time leave detectable signals in the matter power spectrum  allowing for a verification of the proposal.

\section{Cosmological Results and MCMC implementation }\label{sec.results}

We consider here two models, the first one is simple  the standard  \LCDMd model while our second model corresponds to an extension to \LCDM,  where we add  extra relativistic energy density $\rex (a) \propto 1/a^3$  present only at early times for  a scale factor $a$ smaller than  $a_c$,  corresponding to the transition scale factor,  and  the extra energy density dilutes as  $\rex \propto a^{-6}$  for $a\gg a_c$ and becomes therefore rapidly negligible.  We refer to this later model as \LCDMxd and is motivated by Bound Dark Energy  (BDE)  model \cite{delaMacorra:2018zbk,Almaraz:2018fhb}  and  EDE models \cite{Linder:2008nq, Francis:2008md, Calabrese:2010uf,Appleby:2013upa}.

With this rapid dilution we  avoid a step function transition  at $a_c$ in the evolution of $\rex (a)$.  Clearly $\rex$ dilutes faster than radiation for $a>a_c$ and its contribution becomes rapidly subdominant.  The energy density $\rex$ can also be parametrized  by the number of extra relativistic degrees of freedom $N_{ex}$,   defined in terms of the  the neutrino temperature $T_\nu$ as $\rex=(\pi^2/30) N_{ex}T_\nu^4$.
We implement the \LCDMxd  model in the Boltzmann  code  CAMB   \cite{Lewis:1999bs, Howlett:2012mh, Lewis:2013hha,Lewis:2002ah,Lewis:2019xzd})  and make a full COSMO-MC (Markov Chains)  analysis for \LCDMd and \LCDMxd for  several data sets and we present  the results and conclusions in section (\ref{sec.analysis}). 

Since our main  interest  here is to study the tension between the inferred value of \H  from  early CMB physics   and  late time  local measurements of  \H  we  use  the CMB  (TT, TE, EE+lowE)  data set from Planck 2018 [Pl-18]  \cite{Aghanim:2018eyx}  and  the recent  measurements from \SHES\,  H$_0^R=(74.03 \pm 1.42 $)  at 68\% confidence level by Riess et al  [R19] \citep{Riess:2019cxk}.   We run MCMC  for both models, \LCDM\, and \LCDMx,  and compare the posterior probabilities and  we assess the viability to alleviate the  $H_0$ tension between CMB from Planck [P18] and  local \H measurements [R19]. 
We decided not to use BAO measurements, keeping in mind that BAO is  consistent with high and low values of \H and it is in the context of \LCDM\, that BAO measurements hint for a lower value of \H \cite{Alam:2020sor}. Besides BAO analysis  is strongly impacted by the late time dynamics of dark energy at  low redshifts $z < 5$.  Changes from a dynamical the dark energy are beyond the scoop of this work since we want to concentrate  on the tension between CMB and local \H measurements.

For our analysis we consider besides the recent measurement  H$_0^{R}=74.03 \pm \sH $ with $\sH=1.42$  at 68\% confidence level (we will quote all  values of \H  and $\sH$  in units of \ksm)  two forecasting values of   $\sigma_H$  and we  take these forecasting values  as $\sH=1$   and $\sigma_H=0.5$. 
With these two forecasting values of $\sH$ we impose a "tighter  observational" constraint on  \H from local measurements to  study the impact on the posterior probabilities of \H and other relevant cosmological parameters in \LCDM\, and \LCDM-Nx models, and we asses the price we have to pay on the  "Goodness of Fit" of CMB $\chi^2_{cmb}$ for these two forecasting values of H$_0$. Notice however that these forecasting values,  $\sH=1$  and $\sigma_H=0.5$ , are of the same order as the average value obtained  in
the review  L.Verde, A. Riess and T.True   \cite{Verde:2019ivm}  with  an average value of local measurements  of  \H=$(73.03\pm 0.8)$ \ksm.

 \begin{figure*}[ht]
%\begin{multicols}{2}
\begin{center}
\includegraphics[height=0.65\textheight]{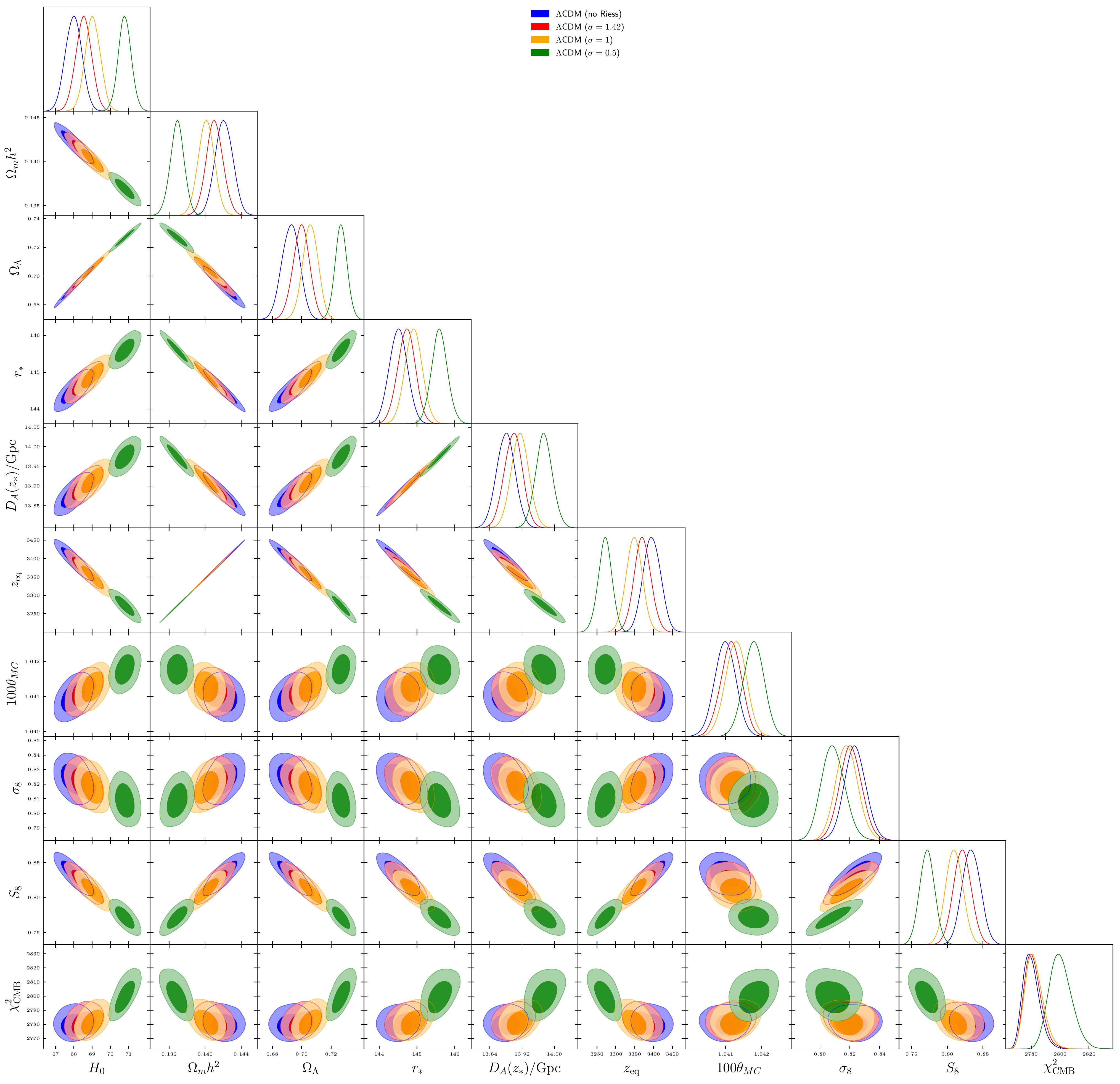}
%\end{multicols}
\caption{We show the marginalized  68\% and 95\% parameters constraint contours  for  \LCDMd  using Planck-2018 TT,TE,EE,lowE  [P18] and  $H_0= (74.03\pm \sigma_H) $\ksm with $\sigma_H=1.42$  [R-19], the forecasting  values  $\sigma_H=1, 0.5$ and \LCDM without Riess data set.}
\label{fig:lcdm}\end{center} \end{figure*}

% sigma_H 1.42
We first consider the MCMC results using CMB data [P-18] and H$_0= (74.02\pm \sH)$ with  $\sigma_H =1.42$ [R-19]. We show  the best fit and the marginalized values at  68\%  confidence level for \LCDM-Nx and \LCDMd for different cosmological parameters  in table \ref{tab:both}. For completeness we also include \LCDM\, without Riess \H data set  (we refer to this case as "No-Riess").
Notice that the value of \H in table \ref{tab:both}  is a slight increased  from \H= $(67.99\pm 0.45)$  in \LCDMd without Riess data [R-19]  to  \H= $(68.54\pm 0.43)$  for \LCDMd and  a value of  \H= $(68.70\pm 0.45)$  in  \LCDM-Nx  considering in these last two cases Riess data [R-19]. The values of \H  correspond to  a mild increase  of  0.81\% and 1.04\%,  in the value of \H  for \LCDM\, and \LCDM-Nx, respectively. These values of \H are still in disagreement with local measurements [R-19]. 
The model \LCDMxd contains extra  radiation  $\Oex (a_c) = 0.063\, (+ 0.146, -0.021)$ with  $N_{ex}= 0.0903\, (+0.28, -0.79)$ at 68\% confidence level.

% Sigma 1, 0.5 
We follow up  our analysis by considering   H$_0^{R}=(74.03 \pm \sigma_H$)  with the  two forecasting values $\sigma_H=1$   and $\sigma_H =0.5$. With these forecasting values on $\sH$  we impose a tighter constraint on the value of \H and this allows to assess the impact on  the posterior probabilities of the cosmological parameters
as well as the Goodness fit for CMB $\chi_{cmb}^2$.   
We  implemented these forecasting  $\sH =1$  and  $\sH =0.5$  in the MCMC analysis for \LCDM\, and \LCDM-Nx models.  We show  the best fit values and posterior probabilities at 68\% c.l.  in table \ref{tab:lcdm} for \LCDMd  and  in table \ref{tab:nx} for \LCDMx.   We find for  \LCDMd  model with the forecasting  $\sH =1$  a value of  \H$=69.04\pm 0.41$ (68\%  c.l.) and a best fit value \H=69.05, while for $\sH =0.5$ we find  \H$=70.79\pm 0.36$  (68\%  c.l.)  and  \H=70.79 for the best fit. While in \LCDMxd\, model we obtain for $\sH =1$  a value   \H$=69.19\pm 0.44$,   with a best fit  \H=69.23 and for  $\sH  =0.5$ we get  \H$=72.99\pm 0.47$ and a best fit   \H=72.83 , respectively. We notice that the impact of a reduced $\sH=0.5$    substantially increases  the value of  \H  in  \LCDMxd   but not in \LCDM\, model. This is no  surprise and is a consequence of the contribution of the extra
relativistic energy density $\rex$ in \LCDMx. 

\begin{figure*}[ht]
%\begin{multicols}{2}
\begin{center} 
\includegraphics[height=0.65\textheight]{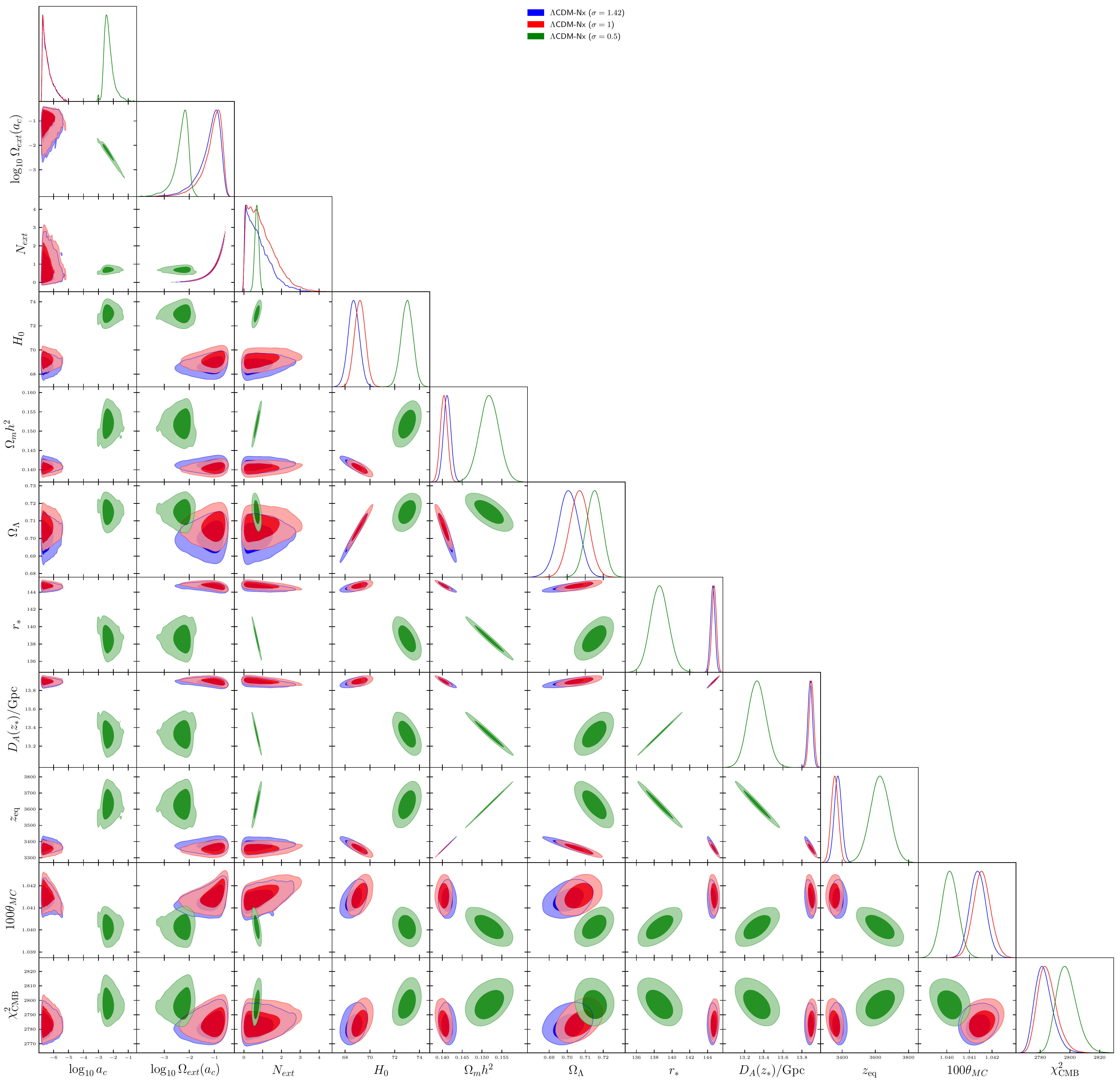}
%\end{multicols}
\caption{We show the marginalized  68\% and 95\% parameters constraint contours  for  \LCDMx\,  using Planck-2018 TT,TE,EE,lowE and  $H_0= (74.03\pm \sigma_H)$ \ksm with   $\sigma_H=1.42$  [R-19]   and  the forecasting  values  $\sigma_H=1, 0.5$. }
\label{fig:nx} \end{center} \end{figure*}

\begin{figure*}[ht]
\begin{center} 
%\begin{multicols}{2}
\includegraphics[height=0.65\textheight]{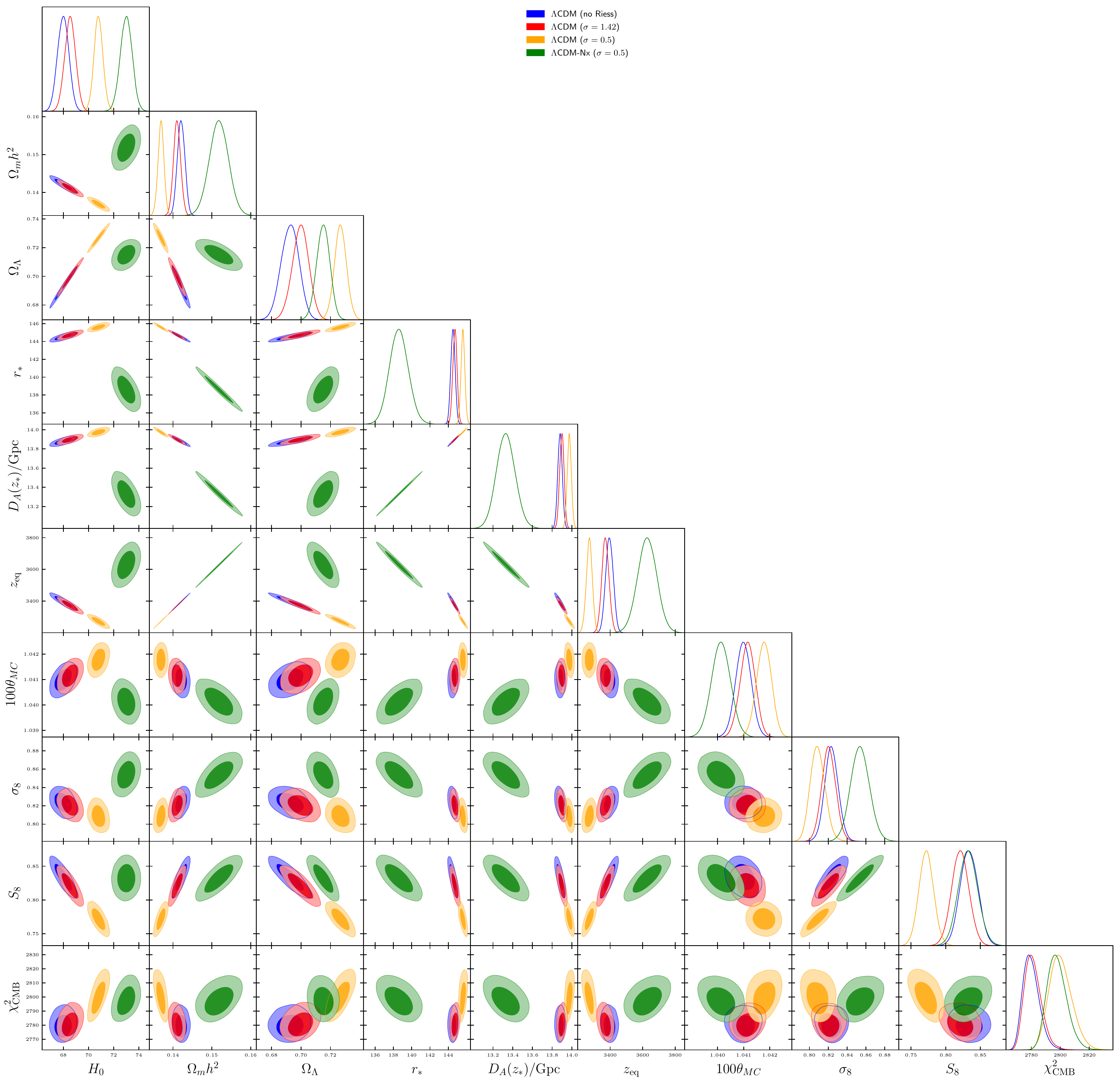}
%\end{multicols}
\caption{We show the marginalized  68\% and 95\% parameters constraint contours  using Planck 2018 TT,TE,EE,lowE 
and  $H_0=(74.03\pm 0.5) $\ksm for \LCDMx\,  and  $H_0=(74.03\pm \sigma_H)$ \ksm\,  with $\sigma_H=1.42$  [R-19]   and  the
forecasting  values   $\sigma_H=0.5$ and  No-Riess  for \LCDM\, models. }
\label{fig:lcdmnx}\end{center} \end{figure*}

We  present  the best fit values and marginalized  68\% and 95\% parameters constraint contours for  different cosmological parameters for  \LCDMd in 
fig.(\ref{fig:lcdm}) and  for \LCDM-Nx in fig.(\ref{fig:nx}).  We find useful to include in a single graph the marginalized  68\% and  95\% parameters constraint contours    \LCDM, $\sH =1.42$, $\sH = 0.5$ and No-Riess  supplemented with  \LCDM-Nx with  $\sH =0.5$  in  fig.(\ref{fig:lcdmnx}). 
This last graph allows for a convenient comparison of the posteriors between  \LCDMd models and \LCDMxd  with $\sH = 0.5$ and the impact on the value of \H and other parameters.

The best fit values for $N_{ex},\, \Oex(a_c)$ and the transition scale factor $a_c$ for the three \LCDMxd cases are:
 $N_{ex}= 0.09$, $\Oex (a_c) = 0.0035$ and   $a_c=(7.1\times10^{-4})$ for \H with  $\sH=1.42$,
$N_{ex}= 0.07$, $\Oex (a_c) = 0.0062$ and   $a_c=(1.5\times10^{-4})$ for \H with $\sH=1$ and
$N_{ex}= 0.61$, $\Oex (a_c) = 0.006$ and   $a_c=(3.48\times10^{-3})$ for \H with $\sH=0.5$.
Notice that the amount  of   $\Oex (a_c)$ remains  of the same order of magintude in all three \LCDMxd  cases while we get an increase of  $N_{ex}$ and $a_c$
by factor of about 10 times  larger  in  \LCDMxd with $\sH=0.5$ compared to  \LCDMxd with $\sH=1.42$ or $\sH=1$.

\begin{table}  \begin{center} \tiny{
\begin{tabular}[t]{ l | rl| rl | rl  }														
\hline  \\ [-5pt]														
	Model	&	$\Lambda$CDM-Nx	&	$H_0 =  (74.03 \pm 1.42) $	&	$\Lambda$CDM	&	$H_0 =  74.03 \pm 1.42 $	&	$\Lambda$CDM	&	H$_0$\, No-Riess	\\
	Parameter	&	Best Fit	&	Sampling	&	Best Fit	&	Sampling	&	Best Fit 	&	Sampling	
	\\ [2pt]\hline \hline \\ [-3pt]														
$	a_c         	$ &  	0.00071	&	$0.407^{+0.105}_{-0.241}\times 10^{-6}$	&	---	&	---	&	---	&	---	\\[3pt]
$	\Omega_{ex}(a_c)	$ &  	0.00353	&	$0.063^{+0.146}_{-0.021}$	&	---	&	---	&	---	&	---	\\[3pt]
$	N_{ex}	$ &  	0.09034	&	$0.81^{+0.22}_{-0.79}      $	&	---	&	---	&	---	& ---		\\ [3pt] 
\hline \\ [-5pt]														
$	H_0	$ &  	69.14	&	$68.70\pm 0.45             $	&	68.557	&	$68.54\pm 0.43             $	&	67.961	&	$67.99\pm 0.45             $	\\
$	\Omega_\Lambda	$ &  	0.702	&	 $0.7007\pm 0.0057          $	&	0.700	&	 $0.6997\pm 0.0056          $	&	0.692	&	$0.6925\pm 0.0060          $	\\
$	\Omega_m	$ &  	0.298	&	$0.2993\pm 0.0057          $	&	0.300	&	$0.3003\pm 0.0056          $	&	0.308	&	$0.3075\pm 0.0060          $	\\
$	\Omega_m h^2	$ &  	0.142	&	$0.14123\pm 0.00093        $	&	0.141	&	$0.14106\pm 0.00092        $	&	0.142	&	$0.14209\pm 0.00096        $	\\
$	\Omega_b h^2	$ &  	0.022	&	$0.02257\pm 0.00014        $	&	0.022	&	$0.02248\pm 0.00013        $	&	0.022	&	$0.02237\pm 0.00013        $	\\
$	z_{eq}	$ &  	3398.84	&	$3375\pm 22                $	&	3370.15	&	 $3371\pm 22                $	&	3397.99	&	$3396\pm 23                $	\\ 
$	ln(10^{10} A_s)	$ &  	3.048	&	$3.049\pm 0.017            $	&	3.047	&	$3.046\pm 0.017            $	&	3.044	&	$3.045\pm 0.016            $	\\
$	n_s	$ &  	0.973	&	$0.9722^{+0.0043}_{-0.0049}$	&	0.970	&	$0.9683\pm 0.0037          $	&	0.966	&	$0.9655\pm 0.0038          $	\\
$	\sigma_8	$ &  	0.826	&	$0.8237\pm 0.0079          $	&	0.821	&	$0.8206\pm 0.0075          $	&	0.824	&	$0.8235\pm 0.0072          $	\\
$	S_8	$ &  	0.823	&	 $0.823\pm 0.013            $	&	0.821	&	$0.821\pm 0.012            $	&	0.835	&	$0.834\pm 0.013            $	\\
$	z_{drag}	$ &  	1089.81	&	$1060.63^{+0.38}_{-0.56}   $	&	1060.12	&	$1060.09\pm 0.28           $	&	1059.93	&	$1059.92\pm 0.28           $	\\
$	r_{drag}	$ &  	146.60		&	$101.11\pm 0.76            $	&	147.35	&	$147.35\pm 0.24            $	&	147.14	&	$147.17\pm 0.24            $	\\
$	z_\star	$ &  	1060.24	&	$1089.95^{+0.26}_{-0.35}   $	&	1089.63	&	$1089.65\pm 0.21           $	&	1089.88	&	$1089.88\pm 0.22           $	\\
$	r_\star	$ &  	143.98&	$144.58\pm 0.25            $	&	144.72	&	$144.72\pm 0.23            $	&	144.48	&	$144.51\pm 0.24            $	\\
$	D_A (r_\star)/\rm{Gpc}	$ &  	13.829	&	 $13.885\pm 0.024           $	&	13.899	&	$13.898\pm 0.022           $	&	13.877	&	$13.880\pm 0.023           $	\\
$	100 \theta\, (z_\star)	$ &  	1.0410	&	$1.04137\pm 0.00036        $	&	1.0411	&	$1.04115\pm 0.00029        $	&	1.0410	&	$1.04099\pm 0.00029        $	\\
 [2pt]\hline \\ [-5pt]														
$	\chi^2_{H0}	$ &  	11.84	&	$14.2\pm 2.4               $	&	14.854	&	 $15.0\pm 2.4               $	&	---	&	---	\\
$	\chi^2_{CMB}	$ &  	2766.24	&	$2782.6\pm 6.1             $	&	2765.69	&	$2781.3\pm 5.8             $	&	2764.35	&	$2780.0\pm 5.7             $	\\
[2pt]\hline 
\end{tabular}}
\caption  {\small We show the best fit, marginalized and 68\% confidence  limits on cosmological parameters for \LCDM-Nx  and \LCDM\,  with Planck-2018 TT,TE,EE-lowE and local $H_0$ R-19 measurements  and \LCDM\, without R-19 (i.e. "No-Riess").
} \label{tab:both}\end{center} \end{table}	

\begin{table} \begin{center}  \tiny{									
\begin{tabular}[t]{  l | rl | rl  }										
\hline  \\ [-5pt]										
	Model	&	$\Lambda$CDM	&	$H_0 =  74.03 \pm 1 $	&	$\Lambda$CDM	&	$H_0 =  74.03 \pm 0.5$	\\
	Parameter 	&	Best Fit 	&	Sampling	&	Best Fit 	&	Sampling	\\ [2pt]\hline \hline \\ [-5pt]	
$	H_0	$ &  	69.050	&	$69.04\pm 0.41             $	&	70.789	&	$70.79\pm 0.36             $	\\
$	\Omega_\Lambda	$ &  	0.706	&	$0.7059\pm 0.0052          $	&	0.727	&	 $0.7268\pm 0.0041          $	\\
$	\Omega_m	$ &  	0.294	&	 $0.2941\pm 0.0052          $	&	0.273	&	$0.2732\pm 0.0041          $	\\
$	\Omega_m h^2	$ &  	0.140	&	 $0.14013\pm 0.00089        $	&	0.137	&	$0.13690\pm 0.00077        $	\\
$	\Omega_b h^2	$ &  	0.023	&	$0.02258\pm 0.00013        $	&	0.023	&	$0.02291\pm 0.00013        $	\\
$	z_{eq}	$ &  	3349.60	&	 $3349\pm 21                $	&	3272.72	&	$3272\pm 18                $	\\
$	ln(10^{10} A_s)	$ &  	3.046	&	$3.046\pm 0.017            $	&	3.049	&	 $3.051^{+0.017}_{-0.019}   $	\\
$	n_s	$ &  	0.972	&	$0.9712\pm 0.0036          $	&	0.981	&	 $0.9805\pm 0.0036          $	\\
$	\sigma_8	$ &  	0.818	&	$0.8180\pm 0.0074          $	&	0.808	&	$0.8089^{+0.0072}_{-0.0082}$	\\
$	S_8	$ &  	0.810	&	$0.810\pm 0.012            $	&	0.772	&	 $0.772\pm 0.011            $	\\
$	z_{drag}	$ &  	1060.31	&	 $1060.24\pm 0.27           $	&	1060.77	&	$1060.73\pm 0.28           $	\\
$	r_{drag}	$ &  	147.48	&	$147.51\pm 0.24            $	&	148.08	&	 $148.11\pm 0.23            $	\\
$	z_\star	$ &  	1089.41	&	$1089.44\pm 0.20           $	&	1088.72	&	$1088.74\pm 0.18           $	\\
$	r_\star	$ &  	144.89	&	 $144.91\pm 0.23            $	&	145.58	&	$145.60\pm 0.21            $	\\
$	D_A (r_\star)/\rm{Gpc}	$ &  	13.913	&	$13.915\pm 0.022           $	&	13.973	&	$13.974\pm 0.021           $	\\
$	100 \theta\,  (z_\star)	$ &  	1.0413	&	 $1.04128\pm 0.00029        $	&	1.0418	&	 $1.04178\pm 0.00028        $	\\
 [2pt]\hline \\ [-5pt]										
$	\chi^2_{H0}	$ &  	24.797	&	$25\pm 4                   $	&	42.027	&	$42\pm 9                   $	\\
$	\chi^2_{CMB}	$ &  	2766.43	&	 $2782.1\pm 6.2             $	&	2784.45	&	$2800.3\pm 7.8             $	\\
[2pt] \hline \end{tabular}} 
\caption  {\small We show the best fit, marginalized and 68\% confidence  limits  for  \LCDM\,  with Planck-2018 TT,TE,EE-lowE and
 $H_0=74.02\pm \sigma_H$ with forecasting value $\sigma_H=1$  and $\sigma_H=0.5$}
\label{tab:lcdm} \end{center} \end{table}

\begin{table} \begin{center} \tiny{									
\begin{tabular}[t]{ l | rl | rl  }										
\hline  \\ [-5pt]										
	Model	&	$\Lambda$CDM-Nx	&	$H_0 =  74.03 \pm 1 $	&	$\Lambda$CDM-Nx	&	$H_0 =  74.03 \pm 0.5$	\\
	Parameter	&	Best Fit 	&	Sampling	&	Best Fit 	&	Sampling 
		\\ [2pt]\hline \hline \\ [-5pt]	
$	a_c          	$ &  	0.00015	&	$0.407^{+0.105}_{-0.245}\times 10{-6}$	&	0.00348	&	$4.898^{+2.247}_{-2.710}\times 10{-3}$	\\[3pt]
$	\Omega_{ex}(a_c)	$ &  	0.00623	&	$0.079^{+0.161}_{-0.023}$	&	0.00603	&	$4.786^{+5.447}_{-1.319}\times 10^{-3}$	\\[3pt]
$	N_{ex}	$ &  	0.07059	&	 $0.99^{+0.28}_{-0.95}      $	&	0.60916	&	 $0.69\pm 0.10              $	\\ [3pt]
 \hline \\ [-1pt]										
$	H_0	$ &  	69.23	&	$69.19\pm 0.44             $	&	72.83	&	$72.99\pm 0.47             $	\\
$	\Omega_\Lambda	$ &  	0.705	&	 $0.7067\pm 0.0053          $	&	0.718	&	$0.7151\pm 0.0045          $	\\
$	\Omega_m	$ &  	0.2949	&	$0.2933\pm 0.0053          $	&	0.2825	&	$0.2849\pm 0.0045          $	\\
$	\Omega_m h^2	$ &  	0.1413	&	$0.14039\pm 0.00090        $	&	0.1499	&	$0.1518\pm 0.0024          $	\\
$	\Omega_b h^2	$ &  	0.0227	&	$0.02267\pm 0.00015        $	&	0.0230	&	 $0.02300\pm 0.00012        $	\\
$	z_{eq}	$ &  	3377.26	&	$3355\pm 22                $	&	3581.23	&	$3628\pm 59                $	\\
$	ln(10^{10} A_s)	$ &  	3.0502	&	$3.051\pm 0.017            $	&	3.0699	&	$3.076^{+0.016}_{-0.018}   $	\\
$	n_s	$ &  	0.9776	&	$0.9753^{+0.0046}_{-0.0052}$	&	0.9911	&	$0.9917\pm 0.0040          $	\\
$	\sigma_8	$ &  	0.8258	&	$0.8220\pm 0.0080          $	&	0.8472	&	$0.854\pm 0.010            $	\\
$	S_8	$ &  	0.8187	&	 $0.813\pm 0.012            $	&	0.8221	&	 $0.832\pm 0.014            $	\\
$	z_{drag}	$ &  	1060.58	&	 $1060.87^{+0.43}_{-0.61}   $	&	1062.30	&	 $1062.51\pm 0.38           $	\\
$	r_{drag}	$ &  	146.92	&	$147.31\pm 0.28            $	&	141.80	&	$141.0\pm 1.0              $	\\
$	z_\star	$ &  	1089.45	&	 $1089.83^{+0.28}_{-0.36}   $	&	1090.31	&	$1090.53\pm 0.32           $	\\
$	r_\star	$ &  	144.36	&	$144.74\pm 0.26            $	&	139.40	&	 $138.6\pm 1.0              $	\\
$	D_A (r_\star)	$ &  	13.860	&	$13.899\pm 0.024           $	&	13.404	&	$13.330\pm 0.095           $	\\
$	100 \theta\, (z_\star)	$ &  	1.0414	&	$1.04155\pm 0.00038        $	&	1.0403	&	$1.04013\pm 0.00036        $	\\
 [2pt]\hline \\ [-5pt]										
$	\chi^2_{H0}	$ &  	23.031	&	$24\pm 4                   $	&	5.71714	&	 $5.2\pm 4.2                $	\\
$	\chi^2_{CMB}	$ &  	2767.06	&	$2784.9\pm 6.4             $	&	2779.81	&	 $2797.9\pm 6.9             $	\\
[2pt]\hline \end{tabular}} 
\caption  {\small We show the best fit, marginalized and 68\% confidence  limits  for  \LCDM-Nx  with Planck-2018 TT,TE,EE-lowE and
 $H_0=74.02\pm \sigma_H$ with forecasting value $\sigma_H=1$  and $\sigma_H=0.5$}
 \label{tab:nx} \end{center}  \end{table}

 \subsection{Analysis}\label{sec.analysis}

Let us now  compare and analyze the results  of the MCMC results in \LCDMd and \LCDMxd models  given in tables  \ref{tab:both},  \ref{tab:lcdm} and  \ref{tab:nx}. Besides these three tables with the best fit and sampling values for different cosmological parameters in  \LCDMd and \LCDMxd  models and the 
corresponding  figures at   68\% and 95\% marginalized parameters constraint contours in  fig.\ref{fig:lcdm} for \LCDM,  fig.\ref{fig:nx} for \LCDM-Nx, and the mixed fig. \ref{fig:lcdmnx}, we find useful to analyze the difference between these cases by determining the relative difference and the percentage difference for some relevant parameters shown in tables \ref{tab:param}  and \ref{tab:diff}, respectively.     

We show in  table  \ref{tab:sigma}  the  discrepancy between the value of  \H=$74.03\pm \sigma_H$,  for the three  different  values of $\sigma_H$ (i.e. $\sigma_H=1.42,1 , 0.5$),  and the posterior probability  of \H $\pm \sigma_s$ with $ \sigma_s$ the 68\% confidence  level for \LCDMd and \LCDMxd from  the MCMC.
The central value of \H of the samplings increases with decreasing $\sH$, while the amplitude of $\sigma_s$ remains nearly constant in all 6 cases ($\sigma_s\sim 0.42$ ). The quantity $\Delta$\H $\equiv$ (74.03 - \H)  corresponds to the distance between  the central value  \H=74.03 from Riess  [R-19] and the central value  \H  from each of the samplings and  we define $\sigma_T\equiv  \sigma_H+\sigma_s$ for each case.
Not surprisingly for smaller values of $\sH$ we obtain a larger H$_0$ and a decrease in $\Delta H_0/\sigma_T$ in \LCDMd and \LCDMxd models.
However, even though the value of \H increases  so does $\chi^2_{H_0}$ in all cases but for \LCDMxd with $\sH=0.5$.  
We obtain in \LCDMd model  $\chi^2_{H_0}=15 \pm 2.4$ for $\sH$ =1.42, $\chi^2_{H_0}=25 \pm 4$  for  $\sH = 1$ and  $\chi^2_{H_0}=42 \pm 9$ for $\sH=0.5$, while in \LCDMxd  we have   $\chi^2_{H_0}=14.2 \pm 2.4$ for $\sH = 1.42 $ and $\chi^2_{H_0}=24 \pm 4$  for  $\sH = 1$ while we have significant reduction in  $\sH = 0.5 $  model obtaining $\chi^2_{H_0}=5.2\pm 4.2$.
Notice that the difference in $\chi^2_{H_0}$ between \LCDMd and \LCDMxd  is small for $\sH=1.42$ and $\sH=1$
however  the impact  from the  forecasting value  $\sH=0.5$  in \LCDMxd has a significant reduction  in $\chi^2_{H_0}$ from   $\chi^2_{H_0}=42 \pm 9$ in \LCDMd to
$\chi^2_{H_0}=5.2\pm 4.2$ in \LCDMx. We  remark  that only   \LCDMxd with $\sH=0.5$   has  an $\Delta H_0/\sigma_T$ smaller than one, clearly showing the impact of the reduced $\sH$.

 \begin{table}  \begin{center} \tiny{														
\begin{tabular}[t] { c | ccc | ccc}														
\hline  \\ [-5pt]	Model	&  	$\Lambda$CDM 	&  	$\Lambda$CDM 	&  	$\Lambda$CDM 	&  	$\Lambda$CDM-Nx 	&  	$\Lambda$CDM-Nx 	&  	$\Lambda$CDM-Nx 	\\
	 $H_0=74.03  \pm \sigma_H$  	&  	$\sigma_H=1.42$	&  	$\sigma_H=1$	&  	$\sigma_H=0.5$	&  	$\sigma_H=1.42$	&  	$\sigma_H=1$	&  	$\sigma_H=0.5$	\\
[2pt] \hline \hline  \\ [-5pt]	$ H_0  \pm \sigma_s $	&  	$68.54\pm 0.43  $	&  	$69.04\pm 0.41  $	&  	$70.79\pm 0.36   $	&  	$68.70\pm 0.45   $	&  	$69.19\pm 0.44   $	&  	$72.99\pm 0.47 $	\\
$\sigma_T=\sigma_H + \sigma_s$	&  	$1.42 +0.43$ 	&  	$1+0.41$	&  	$0.5 +0.36$	&  	$1.42 +0.45$ 	&  	$1+0.44$	&  	$0.5 +0.36$	\\
	$\Delta H_0/\sigma_T$	&  	2.968	&  	2.697	&  	1.820	&  	2.850	&  	2.602	&  	0.550	\\
[2pt] \hline  \\ [-5pt]														
	$\chi^2_{H_0}$	&  	 $15.0\pm 2.4   $	&  	$25\pm 4         $	&  	$42\pm 9   $	&  	$14.2\pm 2.4    $	&  	$24\pm 4    $	&  	 $5.2\pm 4.2   $	\\
%	$\chi^2_{cmb}$	&  	$2781.3\pm 5.8     $	&  	 $2782.1\pm 6.2    $	&  	$2800.3\pm 7.8  $	&  	$2782.6\pm 6.1    $	&  	$2784.9\pm 6.4    $	&  	 $2797.9\pm 6.9    $	\\
[2pt]\hline \end{tabular} }
 \caption  {\small We show  the  central value H$_0$ and the 68\% confidence level  ($\sigma_s$) of  the MCMC samplings  using Planck-2018 and  H$_0=74.02\pm\sigma_H$ with different values of $\sigma_H$ in \LCDM\, and \LCDM-Nx models.   The value $\sigma_H=1.42$ corresponds to local measurements (R-19) while $\sigma_H=1$  and  $\sigma_H=0.5$  the  two  forecasting values of local $H_0$ measurements, while
 $\sigma_s$ corresponds to the sampling  margin at 68\% confidence level. We see that the central value of  $H_0$   increases with decreasing $\sigma_H$ while  the  distance in $\Delta H_0/\sigma_T$ becomes smaller  with $\sigma_T=\sigma_H + \sigma_s$. The reduction is  far more prominent in \LCDM-Nx than in \LCDM.  Finally we show in the last two lines the $\chi^2$ for  $H_0$ and CMB sampling with the different data sets.}
 \label{tab:sigma} \end{center} \end{table}	
\begin{table}  \begin{center}{\tiny{
\begin{tabular}[t]{  l  |  l | cccccccccc }																						
\hline  \\ [-5pt]	Model &   	$100 \theta\,(z_\star) $	&  	$D_A (r_\star)$	&  	$r_\star$	&  	H$_0$	&  	$\Omega_m h^2$	&  	$z_{eq}$	&  	$\sigma_8$	&  	$S_8$	&  	
$\chi^2_{H_0}$	&  	$\chi^2_{cmb}$	\\ 
	$\Lambda$CDM-Nx   $\sigma_H =0.5$  & 	1.04027	&  	13.404	&  	139.40	&  	72.83	&  	0.1499	&  	3581.23	&  	0.847	&  	0.822	&  	5.72	&  	2779.81	\\
[2pt] \hline   \hline \\ [-5pt]	$\Lambda$CDM   $\sigma_H =0.5$ &  	0.146	&  	4.239	&  	4.429	&  	-2.809	&  	-8.612	&  	-8.615	&  	-4.591	&  	-6.156	&  	635.111	&  	0.167	\\
	$\Lambda$CDM  $\sigma_H =1.42$ &  	0.082	&  	3.688	&  	3.815	&  	-5.873	&  	-5.892	&  	-5.894	&  	-3.062	&  	-0.094	&  	159.815	&  	-0.508	\\
	 $\Lambda$CDM   No-Riess &  0.068	&  	3.525	&  	3.639	&  	-6.691	&  	-5.115	&  	-5.117	&  	-2.748	&  	1.524	&  	---	&  	-0.556	\\
[2pt]\hline   \end{tabular} }}
\caption {\small We show  in the 2nd line the best fit values for 	$\Lambda$CDM-Nx   with \H=$(1.42 \pm 0.5)$\ksm  and  we present 
the relative  percent difference  $\Delta_{RPD} P \equiv 100\, (P_\Lambda - P_{Nx})/P_{Nx}$ for different parameters between \LCDM-Nx (with $\sigma_H=0.5$) and \LCDM\, models for different values of $\sigma_H=0.5,1.42$ and No-Riess. } 
\label{tab:param} \end{center} \end{table}

\begin{table} \begin{center} \tiny{																
\begin{tabular}[t] { l | r | rr| rrrr }																	
\hline  \\ [-5pt]																												
	Best Fit	Models 	&	$\Lambda$CDM-Nx	&	$\Lambda$CDM-Nx	&	$\Lambda$CDM-Nx	&	$\Lambda$CDM	&	$\Lambda$CDM	&	$\Lambda$CDM	&	$\Lambda$CDM	\\
		 $H_0 =  74.03 \pm \sigma_H$	&	  $\sigma_H = 0.5$	&	$\sigma_H =1.42$	&	$\sigma_H = 1$	&	No-Riess	&	$\sigma_H = 1.42$ 	&	$\sigma_H = 1$ 	&	$\sigma_H = 0.5$ 	\\
[3pt] \hline \hline  \\ [-5pt]																	
	$	H_0 $ &	72.83	&	69.14	&	69.23	&	67.96	&	68.56	&	69.05	&	70.79	\\
	$	(H_R-H_0)/1.42	$ &	0.83	&	3.43	&	3.37	&	4.27	&	3.85	&	3.50	&	2.28	\\
[3pt] \hline     \\ [-5pt]		Parameter	&	$\Lambda$CDM-Nx	&	 \% Diff.	&	 \% Diff.	&	 \% Diff.	&	 \% Diff.	&	 \% Diff.	&	 \% Diff.	\\
[3pt] \hline\\ [-5pt]	$	H_0	$ &	72.83	&	1.30	&	1.27	&	1.73	&	1.51	&	1.33	&	0.71	\\
	$	\Omega_\Lambda	$ &	0.72	&	0.53	&	0.43	&	0.90	&	0.62	&	0.40	&	-0.32	\\
	$	\Omega_m	$ &	0.28	&	-1.29	&	-1.07	&	-2.15	&	-1.51	&	-1.00	&	0.83	\\
	$	\Omega_m h^2	$ &	0.15	&	1.31	&	1.47	&	1.31	&	-1.51	&	1.67	&	2.25	\\
	$	\Omega_b h^2	$ &	0.02	&	0.55	&	0.33	&	0.66	&	0.54	&	0.41	&	0.07	\\
	$	z_{eq}	$ &	3581.23	&	1.31	&	1.47	&	1.31	&	1.52	&	1.67	&	2.25	\\
	$	\sigma_8	$ &	0.85	&	0.63	&	0.64	&	0.70	&	0.78	&	0.87	&	1.17	\\
	$	S_8	$ &	0.82	&	-0.01	&	0.10	&	-0.38	&	0.02	&	0.37	&	1.59	\\
	$	z_{drag}	$ &	1062.30	&	-0.64	&	0.04	&	0.06	&	0.05	&	0.05	&	0.04	\\
	$	r_{drag}	$ &	141.80	&	-0.83	&	-0.89	&	-0.92	&	-0.96	&	-0.98	&	-1.08	\\
       $	z_\star	$ &	1090.31	&	0.70	&	0.02	&	0.01	&	0.02	&	0.02	&	0.04	\\
	$	r_\star	$ &	139.40	&	-0.81	&	-0.87	&	-0.89	&	-0.94	&	-0.96	&	-1.08	\\
	$	D_A (r_\star)	$ &	13.40	&	-0.78	&	-0.84	&	-0.87	&	-0.91	&	-0.93	&	-1.04	\\
	$	100\, \theta  (z_\star)	$ &	1.04027	&	-0.02	&	-0.03	&	-0.02	&	-0.02	&	-0.02	&	-0.04	\\
[2pt] \hline  \\ [-5pt]	$	\chi^2_{H_0}	$ &	5.72	&	-17.44	&	-30.11	& ---	 &	-22.21	&	-31.26	&	-38.03	\\
	$	\chi^2_{cmb}	$ &	2779.81	&	0.122	&	0.11	&	0.14	&	0.13	&	0.12	&	-0.04	\\
[2pt]\hline \end{tabular} } 
\caption {\small We show the percentage difference  $\Delta P  \equiv 100\, (P_{Nx} -P_\Lambda )/[(P_{Nx}+P_\Lambda )/2] $ of different parameters between \LCDM-Nx (with $\sigma_H=0.5$) and  the different \LCDMxd and  \LCDMd cases with \H=74.03 $\pm\, \sigma_H$  and  No-Riess.  } 
\label{tab:diff} \end{center}  \end{table}

In order to assess  the impact of the reduced  forecasting value $\sH=0.5$  in \LCDMxd\,  on different cosmological parameters  we compare the results from \LCDMxd 
with $\sH=0.5$  and  \LCDMd\,  with $\sH=1.42, \sH=0.5$ and No-Riess  in  table \ref{tab:param}  and we determine the relative percent difference between  \LCDMxd 
with $\sH=0.5$  with \LCDMd  for several parameters and we  show  in table \ref{tab:diff}  the percentage  difference of several parameters between \LCDMxd 
with $\sH=0.5$   and  \LCDMxd  with $\sH=1$ and $\sH=1.42 $ as well as \LCDMd  with $\sH=0.5,\sH=0.5=1, \sH=1.42$ and No-Riess.  

In table \ref{tab:param} we present the relative percent difference  (RPD)  $\Delta_{RPD} P \equiv 100\, (P_\Lambda - P_{Nx})/P_{Nx}$ between \LCDM-Nx (with $\sigma_H=0.5$) and \LCDM\, models (with $\sigma_H=0.5$, $\sigma_H=01.42$ and No-Riess).  Not  surprisingly the change in $\theta$  is small 
($\Delta_{RPD} \theta  < 0.15\%$) while we get a decrease in $D_A(r_\star)$ and  $r(r_\star)$ of the same order ( $\Delta_{RPD} \sim 4\% $  for both quantities),
while  we have  a $\Delta_{RPD} H_0$  of  6.7\%, 5.9\% and 2.8\% with respect to \LCDMd  (No-Riess, $\sH = 1.42$ and $\sH = 0.5$, respectively).
Furthermore, notice that  \LCDMx\, ($\sH = 0.5$) has a significant reduction in  $\chi^2_{H_0}$  compared to \LCDMd  corresponding to a 
$\Delta_{RPD} \chi^2_{H_0} $ of  $635\%$ vs  \LCDM\, ($\sH = 0.5$) and 159\% vs \LCDM\, ($\sH = 1.42$).  
On the other hand,  \LCDMxd ($\sH = 0.5$)   increases  $\chi^2_{cmb}$  by 0.556\% against  \LCDM\, (No-Riess),
0.508\%  vs \LCDM\, ($\sH = 1.42$)  while it reduces  $\chi^2_{cmb}$   by 0.167\%  vs \LCDM\, ($\sH = 0.5$).

\begin{figure*}[ht]
\begin{center} 
%\begin{multicols}
\includegraphics[width=0.495 \textwidth]{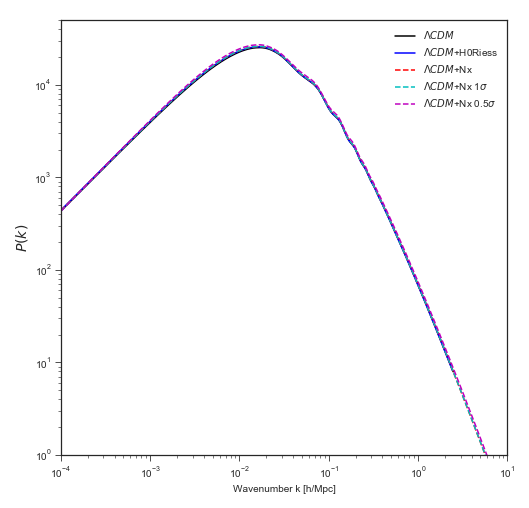}
\includegraphics[width=0.495 \textwidth]{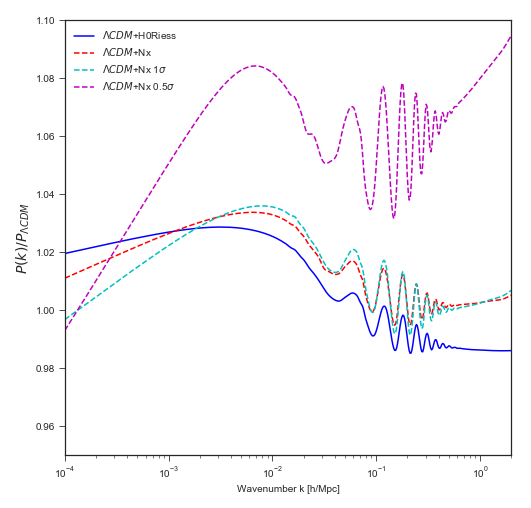}
%\end{multicols}
\caption{We show the matter power spectrum (left panel) for \LCDM and \LCDMxd and  the ratio  (right pannel) \LCDMx/\LCDMd and  \LCDM-NoRiess/\LCDM,
where $1\sigma \equiv $ 1\ksm  and $0.5\sigma \equiv$ 0.5 \ksm }
\label{figMPSS}
\end{center} 
\end{figure*}

\begin{figure*}[ht]
\begin{center}
%\begin{multicols}
\includegraphics[width=0.7\textwidth ]{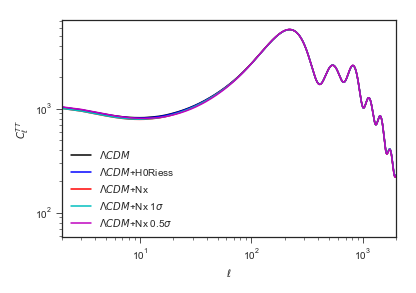}
\includegraphics[width=0.48 \textwidth ]{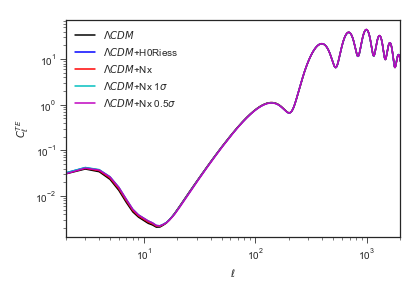}
\includegraphics[width=0.48\textwidth ]{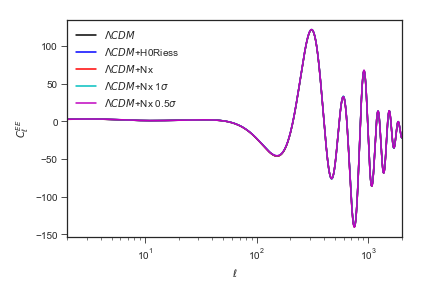}
%\end{multicols}
\caption{\label{powerspectrum} We show the CMB-TT power spectrum from \LCDMx\, and \LCDMd models and their percentage difference, where $1\sigma \equiv $ 1\ksm  and $0.5\sigma \equiv$ 0.5 \ksm.}
\label{figCMBPS}
\end{center} 
\end{figure*}

In table \ref{tab:diff}  we  show  in the two first lines the value of \H and the distance  between obtained Best Fit values of \H for the
different cases and the central value of \H=74.03 divided by the quoted observational error $\sH=1.42$ in [R19].
Notice that only  \LCDMxd (with $\sH=0.5$)  with a value of \H=72.83 has a value below  one $\sH=1.42$
while in all other cases the distance is above two $\sH=1.42$. We also show in  table \ref{tab:diff} 
the percentage  difference $\Delta P  \equiv 100\, (P_{Nx} -P_\Lambda )/[(P_{Nx}+P_\Lambda )/2] $ 
between   \LCDM-Nx ($\sH=0.5$) against  the different cases consider here, i.e. \LCDM-Nx ($\sH = 1.42$ and $\sH=1$) 
and \LCDMd ($\sH = 1.42$, $\sH=1$, $\sH=0.5$ and No-Riess for different cosmological parameters. 
Notice that the change in $\theta (z_\star)$ is quite small and below $0.04$\% while the changes in $r_\star$ and $D_A(\star)$ are of the same order
and  in all cases differ by  approximate 1\%, corresponding to a  percentage change 25 larger than for the $\theta (z_\star)$ quantity.  
We find a significant decrease  in $\chi^2_{H_0}$  for \LCDMxd ($\sH=0.5$) compared to all other 6 cases, with a percentage change of 
$\Delta P (\chi^2_{H_0})= -17.44 $ against \LCDMxd   ($\sH=1.42$)   and  $\Delta P (\chi^2_{H_0})= - 22.21 $ against \LCDMd   ($\sH=1.42$),   and up to an
$\Delta P (\chi^2_{H_0}) =-38.03 $ against \LCDMxd    ($\sH=0.5 $).    On the other hand we get  an increase in percentage  difference   $\chi^2_{cmb}$ 
of at most  0.14\%  for H$_0=74.03\pm 05$  with respect to \LCDM.

\subsection{Matter Power Spectrum and CMB Power Spectrum }\label{sec.MPS}

Here we will show the impact in the matter power spectrum of a rapid diluted energy density given by  $\Oex(a_c)$  at  $a_c$ with  and at a mode $k_c\equiv a_cH_c$ with $H_c\equiv H(a_c)$.  As shown in section \ref{sec.PS} a  rapid diluted energy density generates a bump in the ratio of matter power spectrum between \LCDMxd and  \LCDMd 
model \cite{delaMacorra:2020zqv} and observed in \cite{delaMacorra:2018zbk,Almaraz:2018fhb},  \cite{Calabrese:2011hg,Calabrese:2010uf,Samsing:2012qx,Jaber:2017bpx,Jaber:2016ucq,Jaber:2019opg, Devi:2019hhd} and \cite{Klypin:2020tud,Knox:2019rjx,Francis:2008md}. We show in fig.(\ref{figMPSS}) the matter power spectrum, on the left hand side we plot   \LCDMd with and without Riess data [R-19] and the three \LCDMxd ($\sH=1.42$, $\sH=1$, $\sH=0.5$)   cases.  On the right hand side we show the ratio of \LCDMx/\LCDMd and \LCDM-NoRiess/\LCDM. Notice that for \LCDMxd with $\sH=0.5$ we find an increase in power of about 6\% for modes $10^{-3}< k< 1$ in h/Mpc units while for the other \LCDMd models  ($\sH=1.42$ and $\sH=1$)  the difference is below 2\%. In fig.(\ref{figCMBPS}) we show  CMB power spectrum for all five models described above, top panel corresponds to $C_l^{TT}$ and left bottom panel
 $C_l^{TE}$  and $C_L^{EE}$   right bottom panel.

\section{Conclusions}\label{sec:conclusions}

We have studied  possible solutions to the increasing  \H tension between local \H   and Planck CMB measurements in the context of \LCDMd model. 
Recent local measurements  \H estimate a value of H$_0 = 74.03\pm 1.42$ \ksm  \cite{Riess:2019cxk}  with a  reported average value for different local measurements  of \H=$(73.03\pm 0.8)$ \ksm \cite{Verde:2019ivm} while  Planck reported value of H$_0 = (67.36 \pm 0.54)$ \ksm \cite{Aghanim:2018eyx}.
The magnitude of the tension between the measurements of early and late time is in the range  4.0$\sigma$ and 5.7$\sigma$ \cite{Verde:2019ivm},
 implying an important miss-understanding in either the systematic errors of the observational analysis or may hint towards new physics beyond the concordance cosmological \LCDMd model. 
Here we take the second point of view and study possible solutions to reduce the tension between local \H measurements  and the  Cosmic Microwave Background Radiation  observed by Planck satellite. 
Here we take the second point of view and study possible solutions to reduce the tension between local \H measurements  and the  Cosmic Microwave Background Radiation  observed by Planck satellite.  
Here we have taken this second point of view and we studied models beyond \LCDM.  To alleviate this discrepancy we added  to \LCDMd  extra relativistic energy density $\rho_{ex}$ present at early times  and 
we allowed for  $\rex$ to dilute rapidly (i.e. as $\rex \propto 1/a^6$)  for a scale factor  larger than  $a_c$ and we named this model \LCDMx. With these two phenomenological parameters  we  analyse \LCDMd and \LCDMxd with CMB data and local \H measurements. However, besides taking H$_0 = (74.03\pm \sH)$\ksm  with $\sH=1.42$ we
also  included  two forecasting  one-$\sigma$ standard deviations values,   $\sH=1$ and  $\sH=0.5$,  to assess the impact of these forecasting  \H measurements on the posteriors probabilities of the different cosmological parameters.
We obtained  for \LCDMxd with the forecasting local measurement  H$_0 = 74.03\pm \sH$\ksm  with $\sH=0.5$  and  Planck-2018 CMB  (TT,TE,EE+lowE) 
a value for the Hubble parameter  H$_0 = 72.99\pm 0.47$\ksm  at 68\% c.l.  with  a best fit \H=72.83 \ksm. The relative difference decrease in $\chi^2_H$
in  this \LCDMxd with respect to  \LCDMd  is  $\Delta_{RPD}=635$  while we obtain a  small increase  $\chi_{cmb}$  of $\Delta_{RPD}=0.167$.
For the best fits we have a reduction  in $(H_R-H_0)/1.42$ from 4.27  in  \LCDMd to 0.83 for \LCDMxd ($sH=0.5$)  and
the prize to pay is an increase in  the percentage  difference of  0.14\% for  the CMB $\chi^2_{cmb}$.
Finally we would like to stress that our phenomenological model \LCDMx, and in particular the $\rex$ and $\ac$ parameters,  may  have a sound derivation from extension of the standard model of particle physics as for example BDE or EDE models. These are exiting times to pursue a deeper understanding of our universe in a epoch of high precision observations such as DESI  and  allows to further constraining the building blocks of particle physics.

\acknowledgments
A. de la Macorra  acknowledges support from Project IN105021 PAPIIT-UNAM, PASPA-DGAPA, UNAM and thanks the University of Barcelona for their hospitality. E. Almaraz acknowledges support of a Postdoctoral scholarship by CONACYT.  We also thank M. Jaber and J. Mastache for useful discussions.

{\small
\bibliographystyle{unsrtnat}
\bibliography{h0tension}
}

\end{document}